\documentclass[letter,apj]{emulateapj-rtx4}
\pdfoutput=1
\usepackage{graphicx}
\usepackage{enumerate}
\usepackage{amssymb, amsmath, amsfonts}
\usepackage{gensymb}
\usepackage{mathtools}
\usepackage{natbib}
\usepackage{color}
\usepackage{hyperref}
\usepackage{ulem}
\usepackage{soul}
\usepackage{lscape}
\begin{document}

\altaffiltext{1}{Department of Physics, University of California, Santa Barbara, Santa Barbara, CA 93106, USA}

\title{The Hipparcos-Gaia Catalog of Accelerations}
\author{Timothy D.~Brandt\altaffilmark{1}
}

\begin{abstract}
This paper presents a cross-calibrated catalog of {\it Hipparcos} and {\it Gaia} astrometry to enable their use in measuring changes in proper motion, i.e., accelerations in the plane of the sky.  The final catalog adopts the reference frame of the second {\it Gaia} data release (DR2) and locally cross-calibrates both the scaled {\it Hipparcos}--{\it Gaia} DR2 positional differences and the {\it Hipparcos} proper motions themselves to this frame.  This gives three nearly independent proper motion measurements per star, with the scaled positional difference usually being the most precise.  We find that a linear combination of the two {\it Hipparcos} reductions is superior to either reduction on its own, and address error inflation for both {\it Hipparcos} and {\it Gaia} DR2.  Our adopted error inflation is additive (in quadrature) for {\it Hipparcos} and multiplicative for {\it Gaia}.  We provide the covariance matrices along with the central epochs of all measurements.  Our final proper motion differences are accurately Gaussian with the appropriate variances, and are suitable for acceleration measurements and orbit fitting.  The catalog is constructed with an eye toward completeness; it contains nearly 98\% of the {\it Hipparcos} stars.  It also includes a handful of spurious entries and a few stars with poor {\it Hipparcos} reductions that the user must vet by hand.  Statistical distributions of accelerations derived from this catalog should be interpreted with caution.
\end{abstract}

\maketitle

\section{Introduction} \label{sec:intro}

The {\it Hipparcos} mission, operating between 1989 and 1993, measured astrometry for over 100,000 stars with unprecedented precision \citep{ESA_1997}.  Its successor, {\it Gaia}, has now measured the astrometry of more than 1 billion stars \citep{Gaia_General_2016,Gaia_Astrometry_2018}.  These missions have added parallaxes and proper motions to the more easily measured positions and radial velocities, providing full phase-space information over a large fraction of the Galaxy.  {\it Hipparcos} and {\it Gaia} have enabled the discovery and mapping of Galactic streams \citep{Koppelman+Helmi+Veljanoski_2018}, of stellar moving groups with ever-expanding membership lists \citep{Zuckerman+Song_2001,Zuckerman+Song+Bessell_2004,Malo+Doyon+Lafreniere_2013,Gagne+Faherty_2018}, measurements of the proper motions of globular clusters and dwarf galaxies \citep{Simon_2018,Gaia_Clusters_2018}, and the mapping of the Galactic potential \citep{Price-Whelan+Johnston_2013,Sanderson+Hartke+Helmi_2017}.  

{\it Hipparcos} scanned the sky from 1989 through 1993, mapping the positions, parallaxes, and proper motions of about 118,000 stars.  Each star was observed $\sim$100 times in a variety of spacecraft orientations.  Each observation constrained the position much better in one direction than in its orthogonal direction.  The five-parameter fits and covariance matrices were derived from fits to this epoch astrometry.  The initial {\it Hipparcos} catalog was released three years after the conclusion of the mission \citep{ESA_1997}.  Another reduction of the raw data, completed in 2007, claimed improvements in precision by up to a factor of 4 over the original catalog \citep{vanLeeuwen_2007}.

{\it Gaia}, launched in 2014, uses similar observational principles to {\it Hipparcos}, but with a larger mirror, large-format CCD mosaic focal plane, and vastly superior precision \citep{Gaia_General_2016}.  With its second data release (DR2), {\it Gaia} has now independently measured astrometry of more than one billion stars \citep{Gaia_General_2018,Gaia_Astrometry_2018}.  This raises the possibility of combining the catalogs.  With their $\sim$24-year time baseline, {\it Hipparcos} and {\it Gaia} probe the astrometric acceleration of stars in an inertial frame.  Several authors have recently used these catalogs to derive dynamical masses of planets and brown dwarfs \citep{Calissendorff+Janson_2018,Snellen+Brown_2018}.  

The combination of {\it Hipparcos} and {\it Gaia}, with its $\sim$24-year baseline, is sensitive to orbital periods as long as several hundred years.  The catalogs together provide three proper motion measurements:
\begin{itemize}
    \item The {\it Hipparcos} proper motions at an epoch near 1991.25;
    \item The {\it Gaia} DR2 proper motions at an epoch near 2015.5; and
    \item The ${\it Gaia} - {\it Hipparcos}$ positional difference divided by the $\sim$24-year time baseline.
\end{itemize}
The time baseline is long enough to make the positional difference the most precise proper motion measurement for most stars.  

In order to use these to search for accelerations or to fit orbits, we must first correct any offsets between the catalogs and ensure that the final uncertainties correctly describe the residuals.  For example, the {\it Gaia} team has measured a frame rotation between the DR2 proper motion and the scaled positional difference. \citep{Gaia_Astrometry_2018}.  Simulations performed by the {\it Gaia} team have also shown that the formal uncertainties likely underestimate the true errors, especially for bright stars \citep{Lindegren+Lammers+Hobbs+etal_2012}, while calibrations of the {\it Gaia} reference frames find that the formal uncertainties underestimate the true errors by $\sim$10\% for faint quasars \citep{Gaia_ICRF_2018}.  Calibrations of the reference frames and formal uncertainties have been missing from much of the recent work on astrometric accelerations; they are the core goal and result of this paper.  

The paper is organized as follows.  Section \ref{sec:crosscal} shows that the catalogs in their published format are unsuited for orbit fitting.  In Section \ref{sec:xmatch}, we begin the task of cross-matching {\it Hipparcos} and {\it Gaia}.  Section \ref{sec:gaia_plx} describes our use of {\it Gaia} parallaxes to improve the other {\it Hipparcos} astrometric parameters, while Section \ref{sec:epoch} describes our computation of the central astrometric epoch in each catalog.  Sections \ref{sec:hip_cal} and \ref{sec:gaia_cal} describe our calibrations of the {\it Hipparcos} and {\it Gaia} proper motions to the scaled positional differences in the two catalogs.  In Section \ref{sec:localfit}, we use a cross-validation sample to demonstrate that a local cross-calibration is superior to a global fit, and use this cross-validation set to optimize a Gaussian process regression.  Section \ref{sec:catalog} describes the structure and content of the resulting catalog. We conclude with Section \ref{sec:conclusions}.

\section{The Necessity of a Cross-Calibration} \label{sec:crosscal}

\begin{figure}
    \centering
    \includegraphics[width=\linewidth]{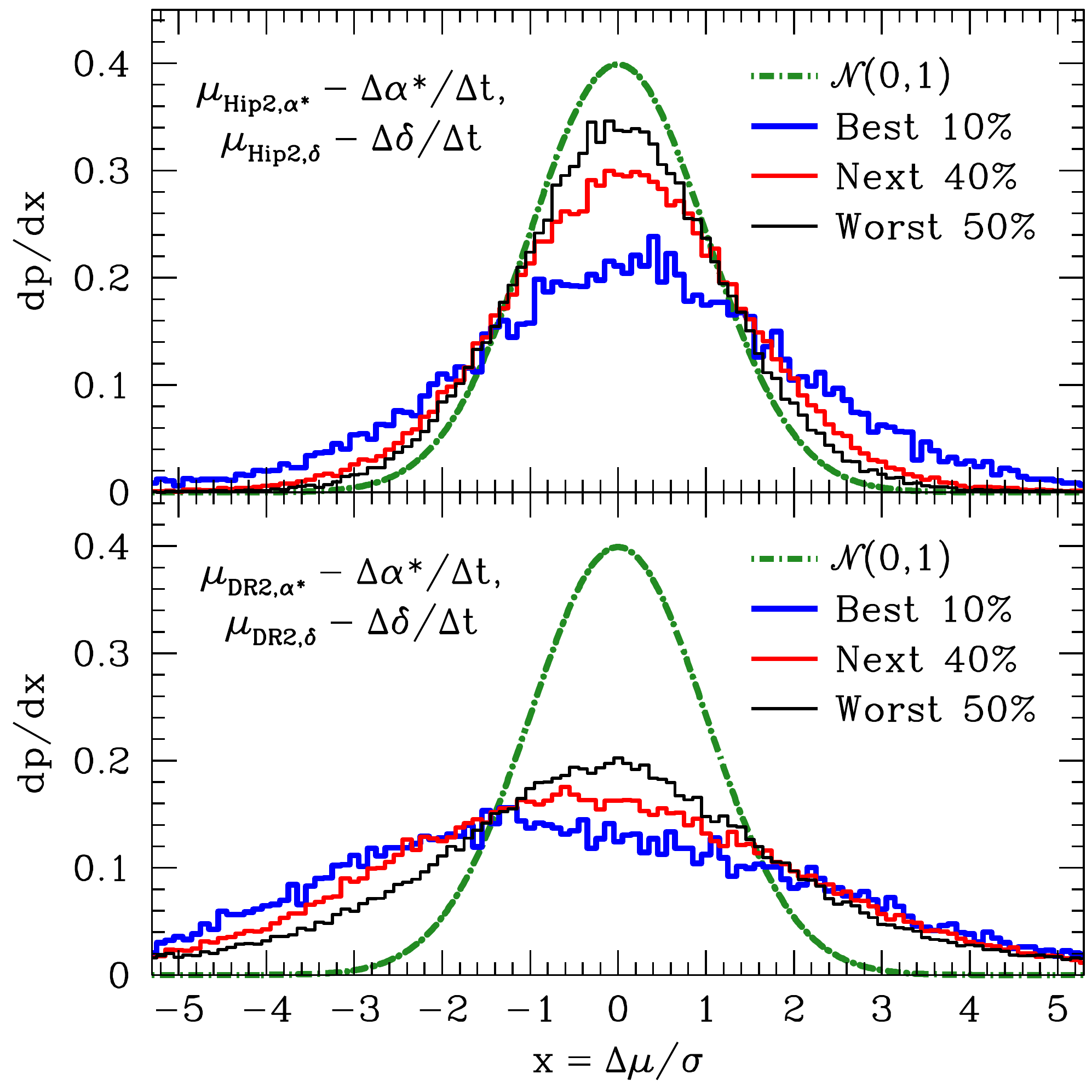}
    \caption{Difference between the {\it Hipparcos}--{\it Gaia} DR2 scaled positional difference and the published proper motion (both in right ascension and declination) in each catalog without applying any cross-calibration.  The blue lines show the stars with the lowest published uncertainties, while the black lines show those with the largest published uncertainties.  Top panel: the {\it Hipparcos} formal uncertainties in the new reduction \citep{vanLeeuwen_2007} do not capture the full distribution of residuals, especially for bright, high-precision stars.  Bottom panel: the poor agreement with {\it Gaia} DR2 \citep{Gaia_Astrometry_2018} is conspicuous because of {\it Gaia}'s exceptional precision.  The wider distributions offset from zero are due to a combination of {\it Hipparcos} frame rotation (which the {\it Gaia} team fit when constructing the TGAS catalog, \citealt{TGAS_Astrometry_2016}), rotation of the {\it Gaia} DR2 reference frame for bright stars (Figure 4 of \citealt{Gaia_Astrometry_2018}), and underestimated uncertainties.}
    \label{fig:baseline_uncertainties}
\end{figure}

The {\it Hipparcos} and {\it Gaia} catalogs taken directly in their published form are unsuited to astrometric orbit fitting.  Both catalogs are calibrated to the International Celestial Reference System \citep[ICRS,][]{Ma+Arias+Eubanks+etal_1998, Fey+Gordon+Jacobs+etal_2015}, but neither one perfectly realizes this reference frame.  Figure \ref{fig:baseline_uncertainties} compares the positional difference between {\it Gaia} DR2 \citep{Gaia_Astrometry_2018} and the new {\it Hipparcos} reduction \citep{vanLeeuwen_2007} with the proper motion in either catalog, normalized by the formal uncertainties.  The figure includes almost 93,000 stars: the subsample of {\it Hipparcos} that we cross-matched to {\it Gaia} DR2 and that have all proper motion residuals within 10\,$\sigma$ of zero (see Section \ref{sec:xmatch}).  The different lines show the 10\% of stars with the lowest proper motion uncertainties, then the next 40\%, and finally the worst 50\%.  

The published uncertainties are based on models of the instruments and have had the uncertainties of the epoch astrometry inflated to achieve formally good fits.  In the case of {\it Gaia} DR2, these additional uncertainties were initially mis-estimated due to the so-called DOF bug; a magnitude-dependent correction is included in the catalog to compensate for this \citep{Gaia_Astrometry_2018}.  For {\it Hipparcos}, these formal uncertainties underestimate the residual scatter, especially for the bright stars with precise measurements.  This has long been anticipated from simulations \citep{Lindegren+Lammers+Hobbs+etal_2012}, and was modeled by \cite{TGAS_Astrometry_2016} for the {\it Tycho-Gaia} Astrometric Solution (TGAS) catalog.  

For {\it Gaia}, the distributions are much broader than the reference Gaussian and are offset from zero.  This is caused by a combination of rotation between the reference frames and by formal uncertainties underestimating the true errors; it is conspicuous thanks to {\it Gaia}'s exceptional precision.  Some of the cross-calibration, including the frame rotation of {\it Hipparcos}, was included in the {\it Tycho}-{\it Gaia} Astrometric Solution \citep[TGAS, ][]{Michalik+Lindegren+Hobbs_2015,TGAS_Astrometry_2016}.  Figure 4 of \cite{Gaia_Astrometry_2018} shows the rotation of the {\it Gaia} DR2 reference frame for the bright stars relative to TGAS, which is the dominant contributor to the offsets in Figure \ref{fig:baseline_uncertainties}.  Adding this frame rotation to the {\it Gaia} proper motions would remove the zero-point offsets and some of the excess scatter.  The frame rotation rates shown in Figure 4 of \cite{Gaia_Astrometry_2018} are consistent with those we derive in Section \ref{sec:gaia_cal} for our global fit.

In the rest of the paper, we cross-match and cross-calibrate the catalogs and uncertainties to make the distributions in Figure \ref{fig:baseline_uncertainties} approximately Gaussian with zero mean and unit variance.  

\section{The Initial Cross-Match} \label{sec:xmatch}

Of the 118,218 stars in the {\it Hipparcos} catalog \citep{ESA_1997}, just 83,034 are present in the {\it Hipparcos}-DR2 best neighbor catalog available on the {\it Gaia} archive \citep{Marrese+Marinoni+Fabrizio+etal_2018}, and 93,635 are present in TGAS \citep{Michalik+Lindegren+Hobbs_2015,TGAS_Astrometry_2016}.  These cross-matched catalogs were constructed to be as robust as possible, and excluded stars that were uncertain matches.  To obtain a higher fraction of stars, including those with significant astrometric accelerations, we perform our own cross-match.  We use generous selection criteria, preferring to include a few false matches in our catalog in order to include all stars showing astrometric accelerations with valid solutions in both catalogs.  

We first perform a coordinate search in the {\it Gaia} archive\footnote{\tt http://archives.esac.esa.int/gaia/} to select potential matches.  For each star in DR2, we propagate its position to 1991.25 assuming the DR2 position and proper motion, and then search within 3$^\prime$ of this propagated position for {\it Gaia} DR2 stars brighter than $G=13$~mag.  This yields nearly three potential matches, on average, for each {\it Hipparcos} star.  It includes potential cross-matches for 117,346 of the {\it Hipparcos} stars, more than 99\% of that catalog.  

We then pare down this initial cross-match catalog by requiring matches in magnitude and parallax within large uncertainties.  The passbands in {\it Hipparcos} and {\it Gaia} differ, making it impossible to directly compare magnitudes.  A very tight positional match of {\it Hipparcos} and {\it Gaia} yields a relatively clean sample for which we can calibrate a color relation.  For nearly all stars, we find
\begin{equation}
   \Delta {\rm mag} = H_p - (0.91B_p + 0.09R_p) < 0.1.
\end{equation}
There is additional scatter from variable and multiple stars.  We use a generous cut, $-0.8 < \Delta{\rm mag} < 0.4$.  We then apply a 10\,$\sigma$ cut on parallax, requiring agreement between the DR2 parallax and a composite {\it Hipparcos} parallax using the combined catalog and error inflation determined later in this paper.  A few hundred variable stars fail the magnitude cut even though they are good matches between the catalogs.  We therefore also include all {\it Gaia} DR2 stars for which propagating their position back to the {\it Hipparcos} epoch (using the DR2 proper motions) yields a positional match within 1$^{\prime\prime}$.  These initial, generous cuts reduce our cross-match catalog to 122,666 DR2 stars near 116,074 {\it Hipparcos} stars.  

We use the cross-match catalog above for our analyses.  To cross-calibrate the catalogs, we select only those stars that are overwhelmingly likely to be valid matches (and unlikely to be accelerating) by requiring all proper motion measurements to be consistent with one another within 10\,$\sigma$; this yields nearly 93,000 stars.  We further use a Gaussian mixture model to handle outliers.  Our final cross-matched catalog, presented at the end of this paper, includes only the best match to each {\it Hipparcos} star, quantified by their agreement in a $\chi^2$ sense.  We discuss this final compilation in Section \ref{sec:catalog}.

\section{Using the {\it Gaia} Parallaxes} \label{sec:gaia_plx}

The goal of the present paper is a catalog of accelerations; we will make no use of the {\it Hipparcos} parallaxes.  However, covariance between the five measured astrometric parameters also means that an improved value for one of them translates into improvements for all.  Reductions of both {\it Hipparcos} and {\it Gaia} report the full covariance matrices.  For the new {\it Hipparcos} reduction \citep{vanLeeuwen_2007}, the procedure to construct the covariance matrix is given in Appendix B of \cite{Michalik+Lindegren+Hobbs+etal_2014}.  A better parallax measurement at the {\it Hipparcos} epoch can improve the other {\it Hipparcos} astrometric parameters.

We begin by propagating the {\it Gaia} parallax to 1991.25 using {\it Gaia}'s measurement of the radial velocity.  This is a negligible correction for nearly all of our stars, but we include it for completeness.  We assume parallax to remain constant for stars without a {\it Gaia} radial velocity.  The corrected parallax, assuming a 24.25-year baseline, is
\begin{align}
    \varpi_{\rm corr} &\approx \varpi_G \left(1 + 2.48 \times 10^{-8} \left(\frac{{\rm RV}_G}{\rm km\,s^{-1}} \right) \left( \frac{\varpi_G}{\rm mas} \right) \right).
\end{align}
The contribution of the radial velocity uncertainty to the corrected parallax is negligible.

We then use the {\it Gaia} parallax to improve the {\it Hipparcos} astrometric parameters.  We denote the {\it Hipparcos} covariance matrix as ${\bf C}_H$, where the parameters are given by the vector 
\begin{equation}
    {\bf p}_H^T =
    \begin{bmatrix}
    \alpha*_H & \delta_H & \varpi_H & \mu_{\alpha*,H} & \mu_{\delta,H}
    \end{bmatrix}.
\end{equation}
Throughout the rest of the paper, we use $\alpha*$ to denote the right ascension times the cosine of the declination, $\alpha \cos \delta$.  The variables $\delta_H$, $\varpi_H$, and $\mu_H$ denote declination, parallax, and proper motion, respectively, all as measured by {\it Hipparcos}.  The measured {\it Gaia} parallax changes these values, but we do not wish to use the {\it Gaia} measurements of the other astrometric parameters.  Adopting the {\it Gaia} covariance matrix pseudo-inverse as
\begin{equation}
    {\bf C}_G^{-1} = 
    \begin{bmatrix}
    0 & 0 & 0 & 0 & 0 \\
    0 & 0 & 0 & 0 & 0 \\
    0 & 0 & 1/\sigma^2_{\varpi,G} & 0 & 0 \\
    0 & 0 & 0 & 0 & 0 \\
    0 & 0 & 0 & 0 & 0 
    \end{bmatrix}
\end{equation}
weights the {\it Gaia} parallax correctly and applies zero weights to the other parameters.  The updated {\it Hipparcos} covariance matrix is then given by
\begin{align}
    {\bf C}^\prime_H = \left( {\bf C}_H^{-1} + {\bf C}_G^{-1} \right)^{-1}
\end{align}
and the updated parameters are given by
\begin{align}
    {\bf p}^\prime_H = {\bf C}^\prime_H \left( {\bf C}_H^{-1} {\bf p}_H + {\bf C}_G^{-1} {\bf p}_G \right).
    \label{eq:updated_params}
\end{align}
To simplify the calculations and avoid round-off error (position coordinates in mas can be $>$10$^8$), we compute $\delta {\bf p}$ by subtracting ${\bf p}_H$ from both sides of Equation \eqref{eq:updated_params}.  This yields
\begin{align}
    \delta {\bf p}_H = {\bf p}^\prime_H - {\bf p}_H = {\bf C}^\prime_H \left( {\bf C}_G^{-1} \left( {\bf p}_G - {\bf p}_H\right) \right).
\end{align}

We apply the parallax corrections separately to the \cite{vanLeeuwen_2007} and \cite{ESA_1997} reductions of the raw {\it Hipparcos} data.  The variances on the {\it Hipparcos} astrometric parameters decrease slightly, by a median amount of $\sim$1\%, after incorporating {\it Gaia} parallaxes.  The median absolute differences between {\it Hipparcos} and {\it Gaia} proper motions also decrease by about 1\%, after adopting the composite {\it Hipparcos} catalog described later.  This is about twice the amount we would expect for well-behaved Gaussian uncertainties.  Our procedure does introduce a very small covariance between the {\it Hipparcos} and {\it Gaia} data, but it is much smaller than the internal covariances between the {\it Hipparcos} parameters.  

In principle we could apply the same logic to improve the {\it Gaia} DR2 proper motions.  However, the {\it Gaia} parallaxes are better than the {\it Hipparcos} parallaxes; there is little value in using {\it Hipparcos} to update the {\it Gaia} astrometry.  Using {\it Hipparcos} anyway (via our composite catalog as described later) gives a negligible change in the median absolute deviation of {\it Hipparcos} and {\it Gaia} proper motions.  A few bright stars have published {\it Hipparcos} uncertainties below published {\it Gaia} uncertainties.  Even for these, the best candidates, adding the {\it Hipparcos} parallax measurements does not improve the residuals.  We therefore choose to keep the DR2 astrometric solutions and covariance matrices as published in the catalog.

\section{The Astrometric Epoch} \label{sec:epoch}

In both {\it Hipparcos} and {\it Gaia}, a star is observed over a series of ``transits,'' each of which measures its position relative to other stars.  A transit generally measures position in one direction much better than in the orthogonal direction.  This gives rise to different effective epochs for the astrometry in right ascension and declination and accounts for some of their covariance.  

{\it Gaia} and {\it Hipparcos} each report their astrometry at a reference epoch: 1991.25 for {\it Hipparcos} and 2015.5 for {\it Gaia} DR2.  These are not, in general, the central epochs of observations of a given target.  Propagating positions away from the central epoch adds uncertainty from the imperfectly measured proper motions.  The characteristic epoch for an object is the one with the smallest positional uncertainty, and this can differ in right ascension and declination.  Denoting $\delta t$ as the difference between the characteristic epoch and the catalog epoch, 
\begin{equation}
    \delta t = t_{\rm best} - t_{\rm catalog},
\end{equation}
we have, e.g., 
\begin{equation}
    \sigma^2_{\alpha}[\delta t_{\alpha}] = \sigma^2_{\alpha}[0] + 2\delta t_{\alpha} {\rm Cov}[\alpha,\mu_{\alpha*}] + (\delta t_{\alpha*})^2 \sigma^2_{\mu_{\alpha*}}.
\end{equation}
Minimizing this yields
\begin{equation}
    \delta t_{\alpha} = -\frac{{\rm Cov}[\alpha,\mu_{\alpha*}]}{\sigma^2_{\mu_{\alpha*}}}.
\end{equation}
The new value of, e.g., right ascension is then
\begin{equation}
    {\alpha}[t_{\rm catalog} + \delta t_{\alpha}] = {\alpha}[t_{\rm catalog}] + \frac{\delta t_{\alpha} \mu_{\alpha*} }{\cos \delta}.
\end{equation}
These values are significantly different from the catalog epochs.  The median absolute deviations are 0.12 years for {\it Hipparcos} in right ascension and 0.15 years for {\it Hipparcos} in declination.  For {\it Gaia}, the median absolute deviations are 0.13 years in right ascension and 0.16 years in declination.  A few stars have central epochs outside a mission's window of observation due to nonstandard astrometric fits \citep{ESA_1997}.  We take $\delta t=0$ and adopt the catalog astrometry as published for {\it Hipparcos} stars with $|\delta t| > 1.3$~years, and for {\it Gaia} stars with $|\delta t| > 0.8$~years.  This applies to just 6 stars in the new {\it Hipparcos} reduction, 27 stars in {\it Gaia} DR2, and 142 stars in the original {\it Hipparcos} catalog.  

After we adjust the values of right ascension and declination, we need to propagate all uncertainties.  Our coordinate transformation matrix is
\begin{equation}
    {\bf M} = 
    \begin{bmatrix}
    1 & 0 & 0 & \delta t_{\alpha} & 0 \\
    0 & 1 & 0 & 0 & \delta t_{\delta} \\
    0 & 0 & 1 & 0 & 0 \\
    0 & 0 & 0 & 1 & 0 \\
    0 & 0 & 0 & 0 & 1
    \end{bmatrix},
\end{equation}
and the covariance matrix becomes
\begin{equation}
    {\bf M} {\bf C} {\bf M}^T.
\end{equation}

Both the position and proper motion may be considered as being measured at time $\delta t$ after the catalog.  The central time for the proper motion may differ slightly from this depending on the skewness of the distribution of transits weighted by their astrometric precision, but this is difficult to compute without access to the full reduction procedure used to produce the catalogs.  Propagating positions to their central epochs results in modest (typically $\lesssim$few percent) improvements in the uncertainties, but it removes correlation between position and proper motion and can be very important when fitting orbits.  Orbit fitting requires knowing both the measured proper motions and the times when these measurements were taken.

\section{Calibrating the {\it Hipparcos} Proper Motions} \label{sec:hip_cal}

For the overwhelming majority of stars in our data set, the most precise proper motion measurement is the positional difference between {\it Hipparcos} and {\it Gaia} divided by the time baseline.  In this section, we use the scaled positional difference as our reference measurement and calibrate the {\it Hipparcos} proper motions to it.  This value is nearly, though not exactly, identical to the proper motion in the {\it Tycho}-{\it Gaia} Astrometric Solution \citep[TGAS, ][]{TGAS_Astrometry_2016}.   

Before calibrating the {\it Hipparcos} astrometry, we must address the fact that there are two reductions of the {\it Hipparcos} data: the original catalog \citep{ESA_1997} and the reduction by \cite{vanLeeuwen_2007} a decade later.  The second reduction has significantly lower formal uncertainties than the original catalog.  We compare the properties of the two catalogs using the great majority of stars that have a negligible change in proper motion over the $\sim$24-year baseline.  For these stars, the differences 
\begin{equation}
    \Delta \mu_{\alpha*} = \frac{\alpha_{\it Gaia} - \alpha_{\it Hip}}{t_{\alpha,{\it Gaia}} - t_{\alpha,{\it Hip}}}\cos \delta - \mu_{\alpha*,\,\it Hip}
    \label{eq:Dmualpha}
\end{equation}
and 
\begin{equation}
    \Delta \mu_\delta = \frac{\delta_{\it Gaia} - \delta_{\it Hip}}{t_{\delta,{\it Gaia}} - t_{\delta,{\it Hip}}} - \mu_{\delta,\,\it Hip},
    \label{eq:Dmudelta}
\end{equation}
show the properties of the noise.  We subtract the {\it Hipparcos} proper motions from the positional differences to adopt a sign convention in which $\delta t > 0$.  

\cite{TGAS_Astrometry_2016} found a small rotation of $\sim$0.24~mas\,yr$^{-1}$ between the {\it Hipparcos} and {\it Gaia} DR1 reference frames using the {\it Hipparcos} positions.  Here, we allow for rotation between the frames defined by the {\it Hipparcos} proper motion and by the scaled position difference.  We fit for the rotations $\omega_X$, $\omega_Y$, and $\omega_Z$ using the Gaussian mixture model described below, finding different values for the \cite{vanLeeuwen_2007} and \cite{ESA_1997} catalogs.  
We fit for these offsets in each case, and subtract them from Equations \eqref{eq:Dmualpha} and \eqref{eq:Dmudelta} to obtain our corrected proper motion residuals.  

\begin{figure}
    \centering
    \includegraphics[width=\linewidth]{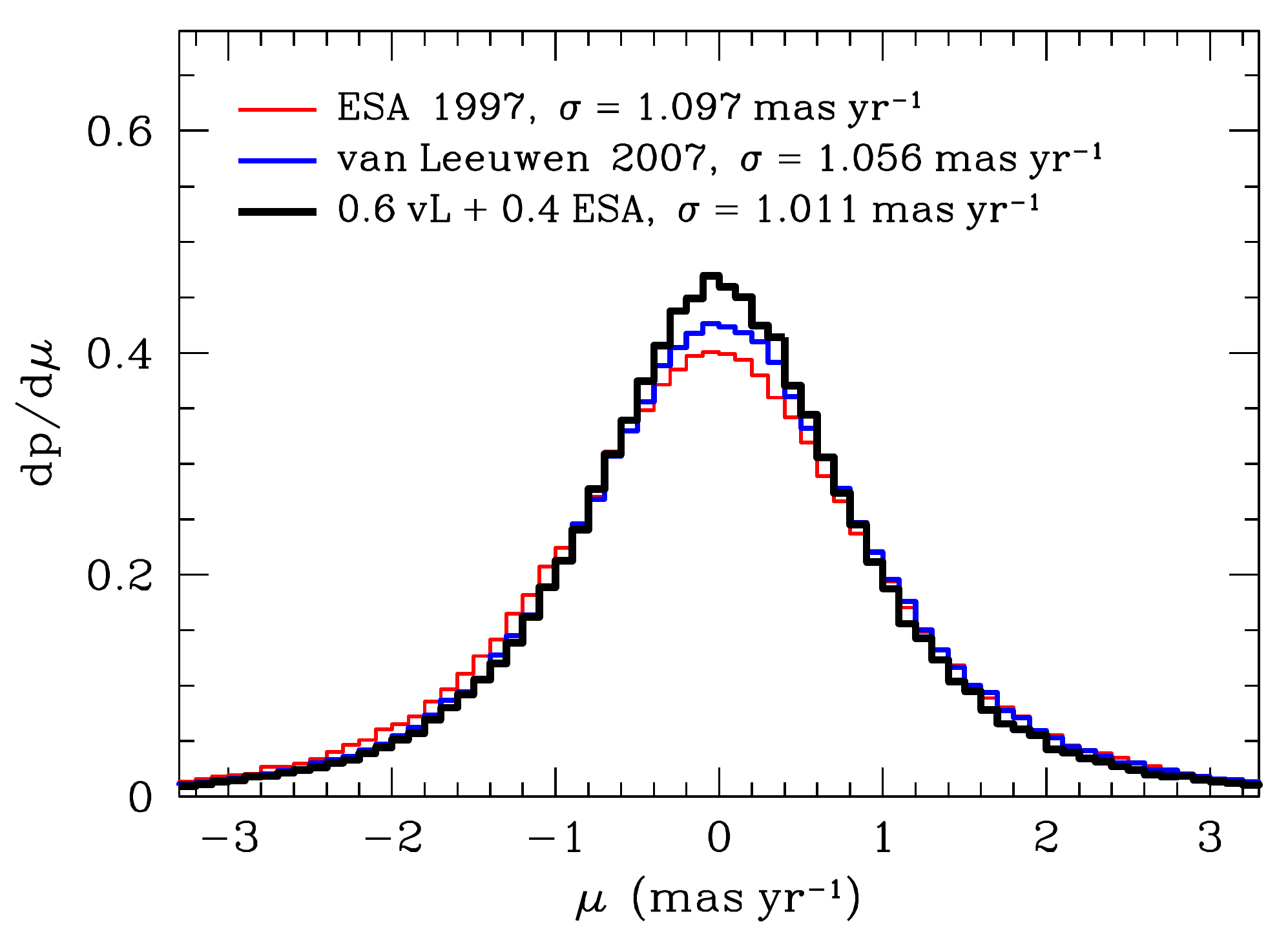}
    \caption{Distribution of residuals of the {\it Hipparcos}--{\it Gaia} scaled positional differences with respect to the proper motion in the original ESA catalog \citep{ESA_1997}, the new {\it Hipparcos} reduction \citep{vanLeeuwen_2007}, and a linear combination of the two catalogs.  Proper motions in right ascension and declination are binned together.  A combination of the two catalogs has lower residuals than either catalog on its own (the figure gives the standard deviations for all points within the plot's limits).  We use a Gaussian mixture model to optimize the weightings of the two catalogs, and to assign a $\sim$150\,$\sigma$ significance to the statement that a 60/40 combination of the {\it Hipparcos} reductions is better than the more precise \cite{vanLeeuwen_2007} reduction on its own.}
    \label{fig:vL_P}
\end{figure}

Figure \ref{fig:vL_P} shows the proper motion residuals, corrected for reference frame rotation, for the original and the new {\it Hipparcos} reductions.  It turns out that a linear combination of the two catalogs is superior to either catalog on its own.  This remains true even when holding the uncertainties fixed to their values in one of the two catalogs.  We address error inflation and the optimal linear combination of the two catalogs using a Gaussian mixture model.

In a Gaussian mixture model \citep[e.g.][]{Ivezic+Connolly+Vanderplas+etal_2014}, each star has a probability $g$ to have its proper motion residual drawn from the claimed error distribution and a probability $1-g$ of it being an outlier and having its proper motion residual drawn from a much broader distribution.  We take our outlier distribution to be a Gaussian with $\sigma=10$~mas\,yr$^{-1}$, much larger than the typical {\it Hipparcos} errors of $\lesssim$1~mas\,yr$^{-1}$.  Marginalizing over $g$ star-by-star adopting a uniform prior\footnote{i.e.~integrating the product of Equation \eqref{eq:gaussmix} and the (uniform) prior over $g$ from $g=0$ to $g=1$} is equivalent to setting $g$ equal to $1/2$.  The results are relatively insensitive to changes in this prior.  

The information needed to reconstruct the full covariance matrix is given in \cite{vanLeeuwen_2007}; Appendix B of \cite{Michalik+Lindegren+Hobbs+etal_2014} explains how to use the data provided to reconstruct the covariance.  The full covariance matrix of the proper motion difference is the sum of this proper motion covariance and the combined {\it Gaia} and {\it Hipparcos} positional covariance matrices divided by the square of the time between the two catalogs.  We assume the errors on the {\it Gaia} position and {\it Hipparcos} proper motion to be uncorrelated and neglect correlations between the {\it Hipparcos} positions and proper motions.  Our propagation of the positions to their central epoch guarantees that these correlations are precisely zero for the components of position and proper motion in the same direction.  For orthogonal directions, e.g.~the covariance between right ascension and proper motion in declination, the positional difference must be divided by the time baseline.  This is squared for the covariance matrix, resulting in a factor $\sim$600 suppression of these terms.  The likelihood of the observed proper motion residual for a given star is then 
\begin{align}
        {\cal L} &= \frac{g}{2\pi \sqrt{\det {\bf C}}} \exp\left[ -\frac{\chi^2}{2} \right] \nonumber \\ 
        &\qquad + \frac{1 - g}{2\pi \sigma^2} \exp\left[ -\left( \frac{\left( \Delta \mu_{\alpha*}\right)^2 + \left(\Delta \mu_\delta \right)^2}{2\sigma^2} \right)\right]
        \label{eq:gaussmix}
\end{align}
with 
\begin{equation}
    \chi^2 = 
    \begin{bmatrix}
    \Delta \mu_{\alpha*} & \Delta \mu_\delta
    \end{bmatrix}
    {\bf C}^{-1}
    \begin{bmatrix}
    \Delta \mu_{\alpha*} \\
    \Delta \mu_\delta
    \end{bmatrix}.
\end{equation}
and $g=1/2$.  We allow for error inflation of the form
\begin{equation}
    {\bf C}^\prime = a^2 {\bf C} + b^2 {\bf I},
    \label{eq:error_inflation}
\end{equation}
i.e., scaling of the {\it Hipparcos} uncertainties by $a$ and the addition of an additional, noncovariant error $b$ in quadrature.  We take the value $b$ to apply identically to the {\it Hipparcos} position and proper motion (with units of either mas or mas\,yr$^{-1}$).  We then fit this model to the full sample of {\it Hipparcos} stars for which our three proper motion measurements agree within 10\,$\sigma$.  Using the \cite{vanLeeuwen_2007} measurements and covariances, we obtain a best-fit $a=1.021$ and $b=0.435$~mas\,yr$^{-1}$.  The best-fit $a$ is very close to 1; with $a$ fixed to unity, the best-fit $b=0.448$~mas\,yr$^{-1}$.

We find the optimal weighting of the \cite{ESA_1997} and \cite{vanLeeuwen_2007} catalogs by varying the relative weights of the measurements, simultaneously fitting for the rotation between the reference frames and the inflation of the errors.  As we decrease the weight on the new {\it Hipparcos} reduction from unity, the best-fit $b$ goes down and the maximum likelihood goes up.  This remains true even if we hold the covariance matrix fixed (before inflation) to the \cite{vanLeeuwen_2007} values.  Holding $a$ fixed at 1, we find that the maximum likelihood and lowest best-fit $b$ both occur for weights of $\sim$60\% for the \cite{vanLeeuwen_2007} astrometry.  With this weighting, and with the pre-inflation covariance matrices fixed to the \cite{vanLeeuwen_2007} values, the best-fit $b = 0.371$\,mas\,yr$^{-1}$, or 77\,$\mu$as\,yr$^{-1}$ lower than for the \cite{vanLeeuwen_2007} catalog itself.  The ratio of the maximum likelihood for the composite catalog to the \cite{vanLeeuwen_2007} catalog is enormous, more than $e^{12000}$.  This corresponds to a Gaussian $\sigma > 150$, decisive evidence in favor of a composite {\it Hipparcos} catalog.

We now address the form of error inflation and the Gaussianity of the residuals.  
Having used the \cite{vanLeeuwen_2007} errors exclusively to this point, we now adopt the weighted sum of the two catalogs' covariance matrices as our baseline uncertainty measurement.  These would be the actual uncertainties if the two catalogs had uncorrelated formal covariances matching the true error distributions.  We find the best-fit values $a$ and $b$ to inflate this covariance matrix using the form given by Equation \eqref{eq:error_inflation}, simultaneously fitting the frame rotation rates and the relative weights of the two catalogs.  After fitting for $a$ and $b$, we divide the stars into three categories: those with the 10\% most precise proper motions, those in the next 40\%, and those in the lower 50\% of {\it Hipparcos} precision.  In each case we show the results of three fits: one with both $a$ and $b$ free, one with $a$ fixed to unity, and one with $b$ fixed to zero.  We compare the normalized distribution of residuals to a Gaussian with zero mean and unit variance.

\begin{figure}
    \centering
    \includegraphics[width=\linewidth]{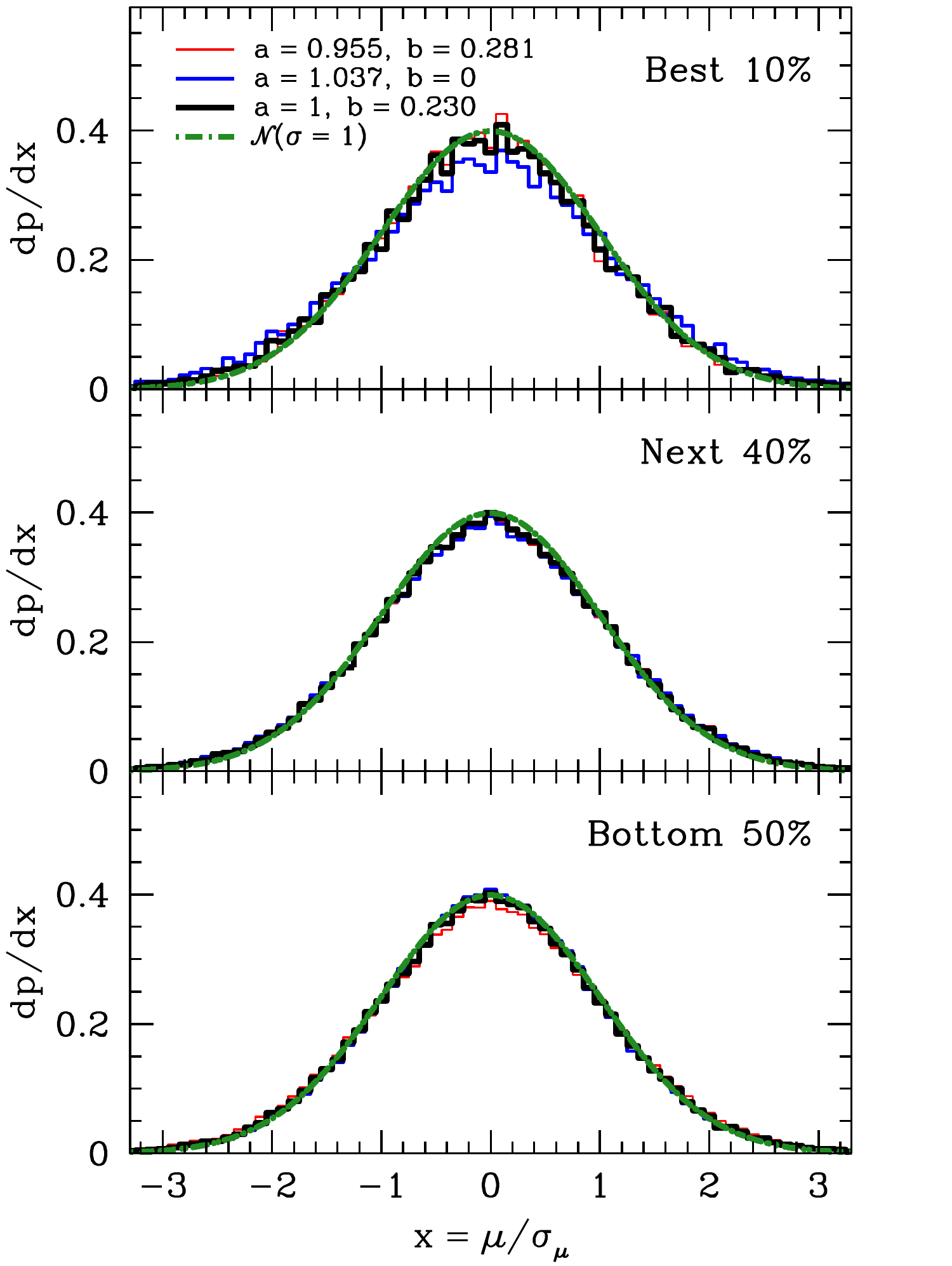}
    \caption{Proper motion residuals in the \cite{vanLeeuwen_2007}/\cite{ESA_1997} composite {\it Hipparcos} catalog, normalized by the composite errors (a weighted sum of the two catalogs' covariance matrices) with the inflation factors (Equation \eqref{eq:error_inflation}) shown.  Multiplying the errors by a constant factor (medium blue curves) underinflates the most precisely measured stars (top panel).  The model with $a$ free (narrow red lines) and with $a$ fixed to unity (thick black lines) are nearly indistinguishable.  We adopt $a=1$, $b=0.230$~mas\,yr$^{-1}$ for our fiducial {\it Hipparcos} model.}
    \label{fig:hip_err}
\end{figure}

Figure \ref{fig:hip_err} shows the results.  Multiplying our errors by a constant factor (medium-width blue histograms) underinflates the errors of the more precise measurements (top panel).  The model with both $a$ and $b$ free (thin red histogram) and the model with $a=1$ (thick black histogram) are nearly indistinguishable by eye.  The Bayesian Information Criterion does favor the more complex model, with a value of 196, corresponding to around 14\,$\sigma$.  However, this would require us to {\it decrease} the uncertainties on the noisiest measurements.  The bottom panel of Figure \ref{fig:hip_err} shows that, for these noisy stars, the model with $a=1$ provides a slightly better match to a Gaussian with unit variance; we adopt $a=1$ as our fiducial choice.  The model with $b$ fixed to zero has a likelihood $e^{504}$ times lower than the model with $a$ fixed to one (corresponding to $\sim$30\,$\sigma$).  

The \cite{vanLeeuwen_2007} reduction has particularly low formal errors for the brightest stars.  We therefore consider the possibility that our weighting of the two catalogs should depend on magnitude, or on the formal uncertainties.  Restricting the analysis to the best 20\% of stars (those with the lowest formal uncertainties) does not change the conclusion above: the best-fit weightings of the catalogs remain very close to 60/40, while the best-fit error inflation factor is slightly higher than when using all of the {\it Hipparcos} stars.  The modeling of the satellite attitude in the \cite{vanLeeuwen_2007} reduction was likely dominated by the bright stars; we speculate that the astrometric solutions and of these stars may have suffered from over-fitting.

\section{Calibrating the {\it Gaia} proper motions} \label{sec:gaia_cal}

We now turn to the {\it Gaia} DR2 proper motions, again calibrating them to the {\it Hipparcos}--{\it Gaia} positional differences (our most precise proper motion measurement).  We fit the same model as for {\it Hipparcos}, allowing for overall rotation of the frame and inflation of the {\it Gaia} uncertainties by both an additive and a multiplicative term.  Our reference proper motion is the difference in position between {\it Gaia} DR2 \citep{Gaia_Astrometry_2018} and our composite {\it Hipparcos} catalog, 60\% \cite{vanLeeuwen_2007} and 40\% \cite{ESA_1997}, divided by the time baseline between the catalogs.  Our uncertainty is the sum of the covariance matrix of {\it Gaia} proper motion and the combined covariance matrix of our composite {\it Hipparcos} position and the {\it Gaia} position divided by the square of the time difference.  As for the {\it Hipparcos} proper motions, we assume the errors on the {\it Hipparcos} position and {\it Gaia} proper motion to be uncorrelated and neglect correlations between the {\it Gaia} positions and proper motions.  Due to our propagation of the positions to their central epoch and the scaling of position differences by the square of the time baseline, these covariances are negligible.

\begin{figure}
    \centering
    \includegraphics[width=\linewidth]{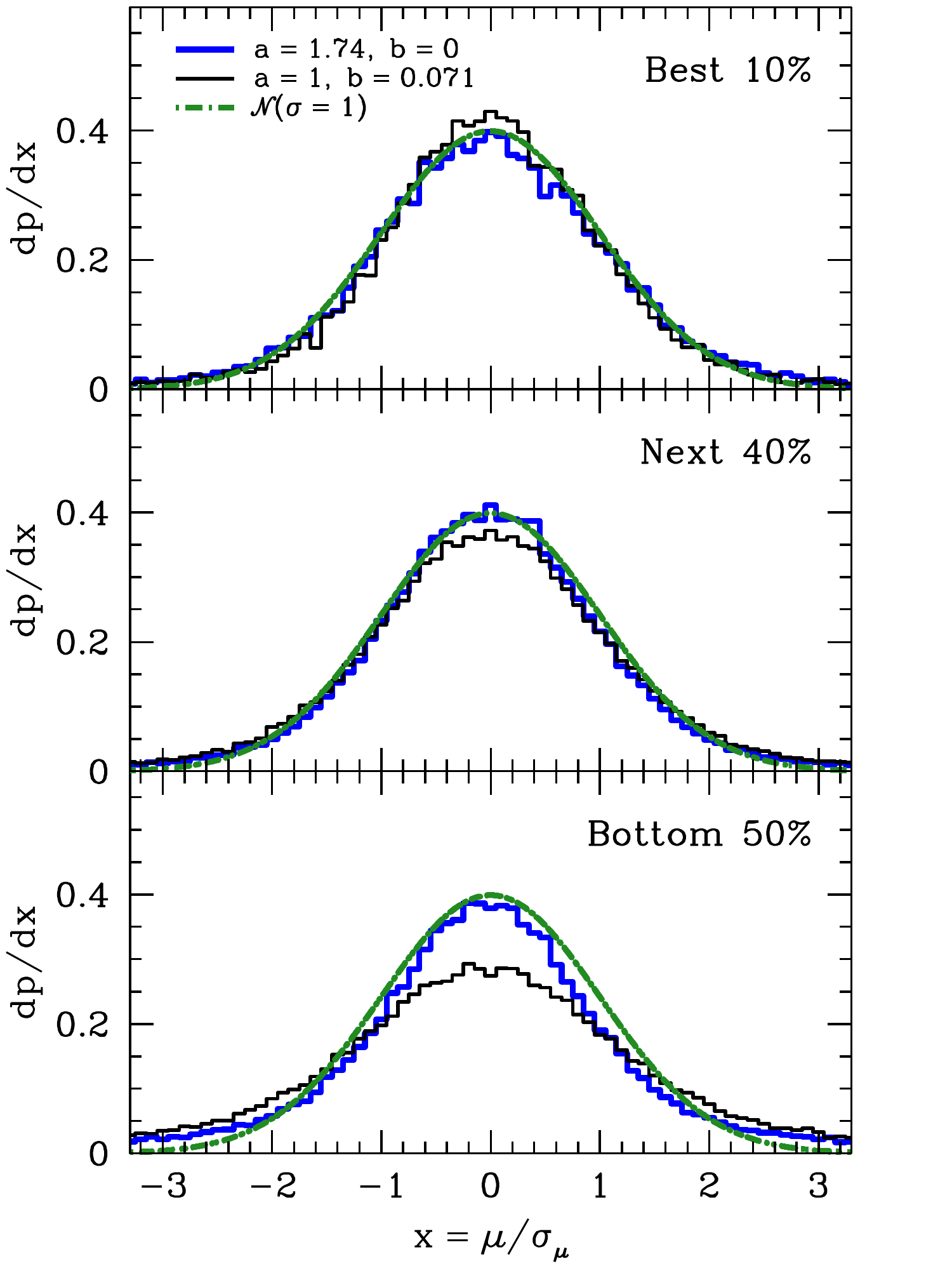}
    \caption{Distribution of the scaled residuals $\Delta \mu_{\alpha*}$, $\Delta \mu_\delta$ under the model with $a$ fixed to unity (thin black line) and with $b$ fixed to zero (thick blue line).  The model with both $a$ and $b$ free has its best fit for $b \approx 0$.  A dot-dashed green line shows a Gaussian with unit variance.  Unlike for {\it Hipparcos}, a constant scaling of the covariance matrix provides the best fit to the {\it Gaia} data.  We obtain a best-fit inflation factor of $\sim$1.74 for the errors, or $1.74^2$ for the covariance matrix.  The distribution of {\it Gaia} residuals has heavier non-Gaussian tails than that for the {\it Hipparcos} residuals, especially for the stars with relatively poor precision.  This may be related to behavior of the pipeline for saturated stars, something that will improve in future data releases, and to the so-called DOF bug.  To fix this bug, the {\it Gaia} team applied a correction factor to the formal uncertainties, but note that the errors for any given star could be incorrect \citep{Gaia_Astrometry_2018}.}
    \label{fig:Gaia_err}
\end{figure}

We simultaneously fit for the frame rotation rate and the error inflation factors.  As for {\it Hipparcos}, we allow for both an additive and a multiplicative error inflation factor, denoting them as $b$ and $a$, respectively.
Figure \ref{fig:Gaia_err} shows our results.  The best-fit model has $a \approx 1.74$ and $b \approx 0$.  The distribution of quasar proper motions \citep{Gaia_ICRF_2018} also shows $b=0$, but has a smaller value of $a \sim 1.1$ for these faint sources.  
The model with $a$ constrained to be 1 (thinner black line) produces a noticeably poorer fit to the data, overinflating precise measurements (top panel) and underinflating noisier measurements (bottom panel).  The likelihood ratio of these two error inflation models is about $e^{11000}$.  

The {\it Gaia} residual proper motions have heavier tails than both a Gaussian of unit variance and the {\it Hipparcos} residual proper motions.  
Some of this may be due to the DOF bug, in which the formal errors were incorrectly calculated in the original data processing and were fixed later \citep{Gaia_Astrometry_2018}.  The {\it Gaia} team notes that while the overall error distribution is meaningful, the formal uncertainties for any given star may be incorrect \citep{Gaia_Astrometry_2018}.  Future data releases will correct this bug and also improve the processing for very bright, saturated stars.

\section{Local vs.~Global Calibrations} \label{sec:localfit}

The preceding analysis shows that we need to fit a minimum of nine parameters to cross-calibrate the {\it Hipparcos} and {\it Gaia} catalogs to the {\it Gaia} DR2 frame:
\begin{itemize}
    \item A rotation vector between DR2 proper motions and the {\it Hipparcos}--{\it Gaia} positional differences (three values);
    \item A rotation vector between {\it Hipparcos} proper motions and the {\it Hipparcos}--{\it Gaia} positional differences (three values);
    \item An error inflation (additive) term for {\it Hipparcos} (one number);
    \item An error inflation (multiplicative) factor for {\it Gaia} (one number); and
    \item The relative weights of the two {\it Hipparcos} reductions (one number).
\end{itemize}
We obtain our final calibration parameters by fitting for all of them simultaneously using our Gaussian mixture model.  The uncertainties on the parameters, estimated using bootstrap resampling, are negligible.  Table \ref{tab:parameterfits} lists the nine global best-fit parameters.

\begin{deluxetable}{lcr}
\tablewidth{0pt}
\tablecaption{{\it Hipparcos}--{\it Gaia} DR2 Best-Fit Global Calibration Parameters}
\tablehead{
    \colhead{Parameter} &
    \colhead{Best-Fit Value} & 
    \colhead{Units}
}  
\startdata
$\omega_X[{\rm DR2}]$ & $-0.081$\tablenotemark{$\dagger$} & mas\,yr$^{-1}$ \\
$\omega_Y[{\rm DR2}]$ & $-0.113$\tablenotemark{$\dagger$} &  mas\,yr$^{-1}$ \\
$\omega_Z[{\rm DR2}]$ & $-0.038$\tablenotemark{$\dagger$} &  mas\,yr$^{-1}$ \\
$\omega_X[{\it Hip}]$ & $-0.098$\tablenotemark{$\dagger\dagger$} & mas\,yr$^{-1}$ \\
$\omega_Y[{\it Hip}]$ & $0.170$\tablenotemark{$\dagger\dagger$} &  mas\,yr$^{-1}$ \\
$\omega_Z[{\it Hip}]$ & $0.089$\tablenotemark{$\dagger\dagger$} &  mas\,yr$^{-1}$ \\
$f$ & 0.599 & \ldots \\
$b[{\it Hip}]$ & 0.226 & mas, mas\,yr$^{-1}$ \\
$a[{\rm DR2}]$ & 1.743 & \ldots 
\enddata
\tablenotetext{$\dagger$}{Compare to Figure 4 of \cite{Gaia_Astrometry_2018}}
\tablenotetext{$\dagger\dagger$}{Compare to Equation (7) of \cite{TGAS_Astrometry_2016}}
\label{tab:parameterfits}
\end{deluxetable}

Having performed a global fit, we now address whether a local, or at least a locally variable, fit is superior.  Local fits to all of the parameters produce higher likelihoods than a single global fit.  However, this procedure runs the risk of overfitting.  We guard against this possibility by holding back 10\% of the {\it Hipparcos} stars (those whose {\it Hipparcos} identification numbers end in zero) as a cross-validation data set.  These stars are similarly distributed across the sky to the full {\it Hipparcos} sample.  We then divide the sky into 920 non-overlapping regions, each containing $\sim$100 stars (of which $\sim$10 are held back for cross-validation).  Each region is defined by a central point and includes all stars that lie closer to this point than to any other region's central point; we choose the locations to give each region almost exactly the same number of non-accelerating stars.

We first fit for the frame rotation rates, holding the other parameters fixed to their global best-fit values.  Local fits cannot constrain rotation about an axis passing through the point.  To avoid problems with unconstrained parameters, we subtract the global best-fit rotation vector {\it before} locally fitting an additional rotation vector.  We then set the rotation about an axis passing through the center of the local fitted region to zero.  As a result, rotation about this point is fixed to its value in the global fit.  After adding the global best-fit rotation back onto these local results, we have 920 measured frame rotation rate vectors.

\begin{figure}
    \centering
    
    \includegraphics[width=\linewidth]{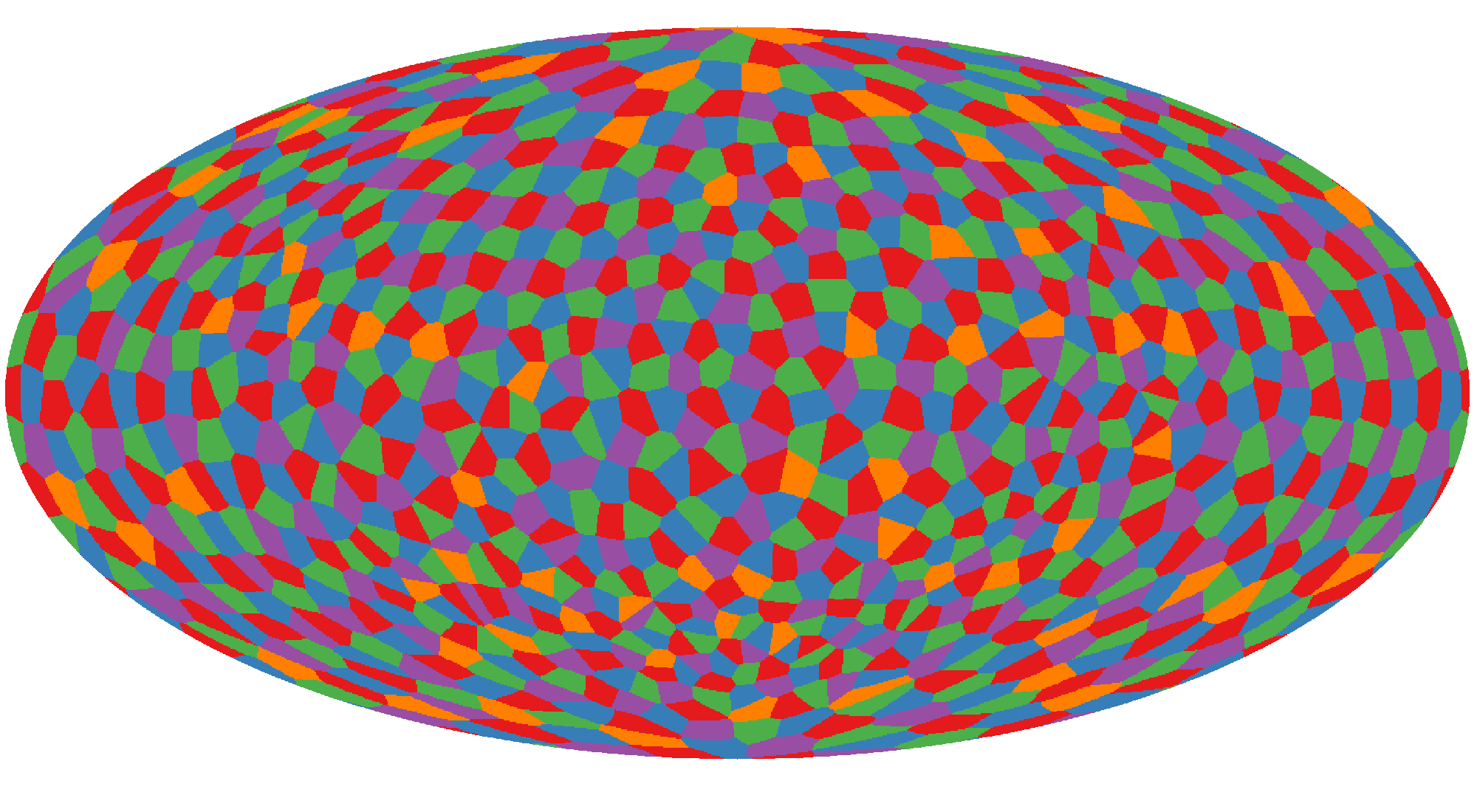}
    \includegraphics[width=\linewidth]{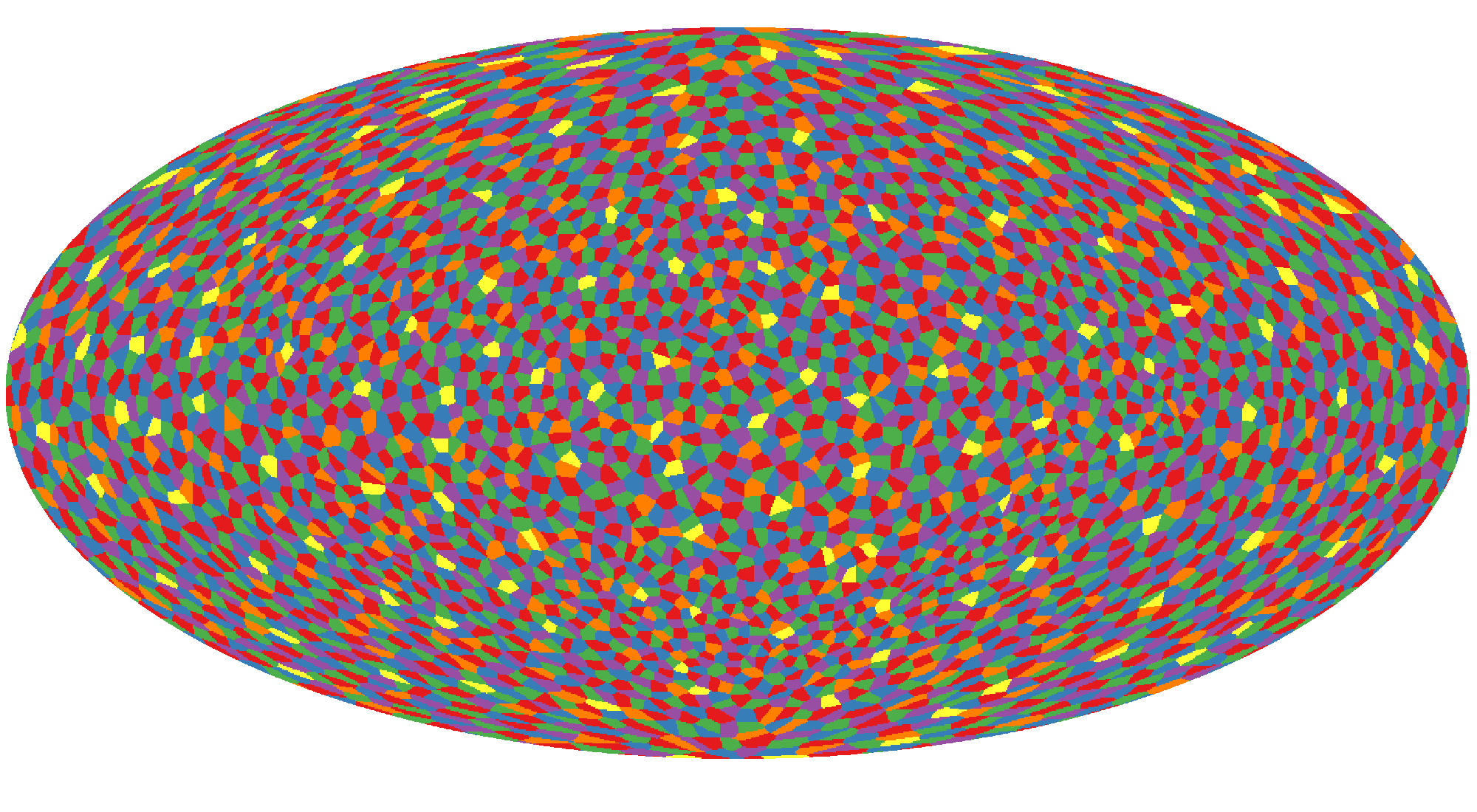}
    \caption{Hammer projections in equatorial coordinates of the regions that we use for our local fits to the {\it Hipparcos}--{\it Gaia} cross-calibration parameters.  The top panel shows the 920 regions that we use for the cross-calibrations of {\it Gaia} to the {\it Hipparcos}--{\it Gaia} position differences.  The bottom panel shows the nearly 5000 regions that we use for the cross-calibration of the {\it Hipparcos} proper motions.  In both cases, we use Gaussian process regression, optimized on a cross-validation sample, to obtain smooth maps.  The point $\alpha = \delta = 0$ lies at the center, east is left, and north is up.  }
    \label{fig:map_coarse_tiled}
\end{figure}

We use our cross-validation sample to compare the performance of the local fits to the rotation rates to that of a global fit.  Our decision to divide the sky into regions each containing 100 stars was arbitrary, so we use Gaussian process regression to obtain a smoothed distribution on the sphere.  Our cross-validation sample provides the data set to optimize the parameters of the regression.  If the global fit is superior to a local fit, the best parameters of the regression will result in very heavy smoothing.  We give each measurement at each point the same variance, equal to the variance of the measurement over the sphere.

The most common covariance function for Gaussian process regression is the squared exponential, i.e.,
\begin{equation}
    \psi_{ij} \propto \exp \left[-d_{ij}^2/(2h^2) \right],
\end{equation}
where $d_{ij}$ is the distance between the points $i$ and $j$ \citep[e.g.][]{Ivezic+Connolly+Vanderplas+etal_2014}.  However, this is not a valid covariance function on the sphere taking $d$ to be either the great-circle distance or the chord distance.  The function is not positive definite: without a large additional diagonal component, the resulting covariance matrix has negative eigenvalues.  We therefore adopt the Mat\'ern class of covariance functions, i.e.,
\begin{equation}
    \psi_{ij} = \sigma^2 2^{\nu - 1}\left(\Gamma[\nu]\right)^{-1} \left(d_{ij}/c\right)^\nu {\cal K}_\nu \left[d_{ij}/c\right]
\end{equation}
where ${\cal K}_\nu$ is the modified Bessel function of the second kind of order $\nu$, $\Gamma$ is the gamma function, and $d_{ij}$ is the great-circle distance.  This function is positive definite on the sphere for $c > 0$ and $0 < \nu \leq 0.5$ \citep{Gneiting_2013}.

We use the same Gaussian mixture model as before to compute the likelihood of our cross-validation sample as a function of $\sigma^2$, $\nu$, and $c$.  For the difference between the {\it Gaia} DR2 proper motions and the scaled positional difference, the likelihood ratio after fitting $\sigma^2$, $\nu$, and $c$ is a highly significant $e^{211}$ in our cross-validation sample.  We then use a finer tiling of the sky than that shown in the top panel of Figure \ref{fig:map_coarse_tiled} to see whether this improves the results.  For the {\it Gaia} proper motions, doing so offers little improvement, from a likelihood ratio of $\sim$$e^{211}$ to $\sim$$e^{220}$ in the cross-validation sample.  It also increases the risk of over-fitting when also constraining the error inflation factor $a$ as we do below.  Figure 13 of \cite{Gaia_Astrometry_2018} suggests that there may be additional correlated errors on a scale of $\sim$1~deg$^2$, but the {\it Hipparcos} source density is insufficient to test this possibility.

For {\it Hipparcos}, the story is different.  Our initial tiling of the sky, with $\sim$100 stars per tile, produces an improvement factor of $e^{127}$ in our cross-calibration data set after optimizing the Gaussian process regression.  This number increases sharply with finer tiling.  With just $\sim$20 stars per tile, or nearly 5000 tiles across the sky (lower panel of Figure \ref{fig:map_coarse_tiled}), a Gaussian process regression fit produces an improvement in the likelihood on the cross-validation sample by a factor of more than $e^{300}$.

\begin{figure*}
    \includegraphics[width=0.5\linewidth]{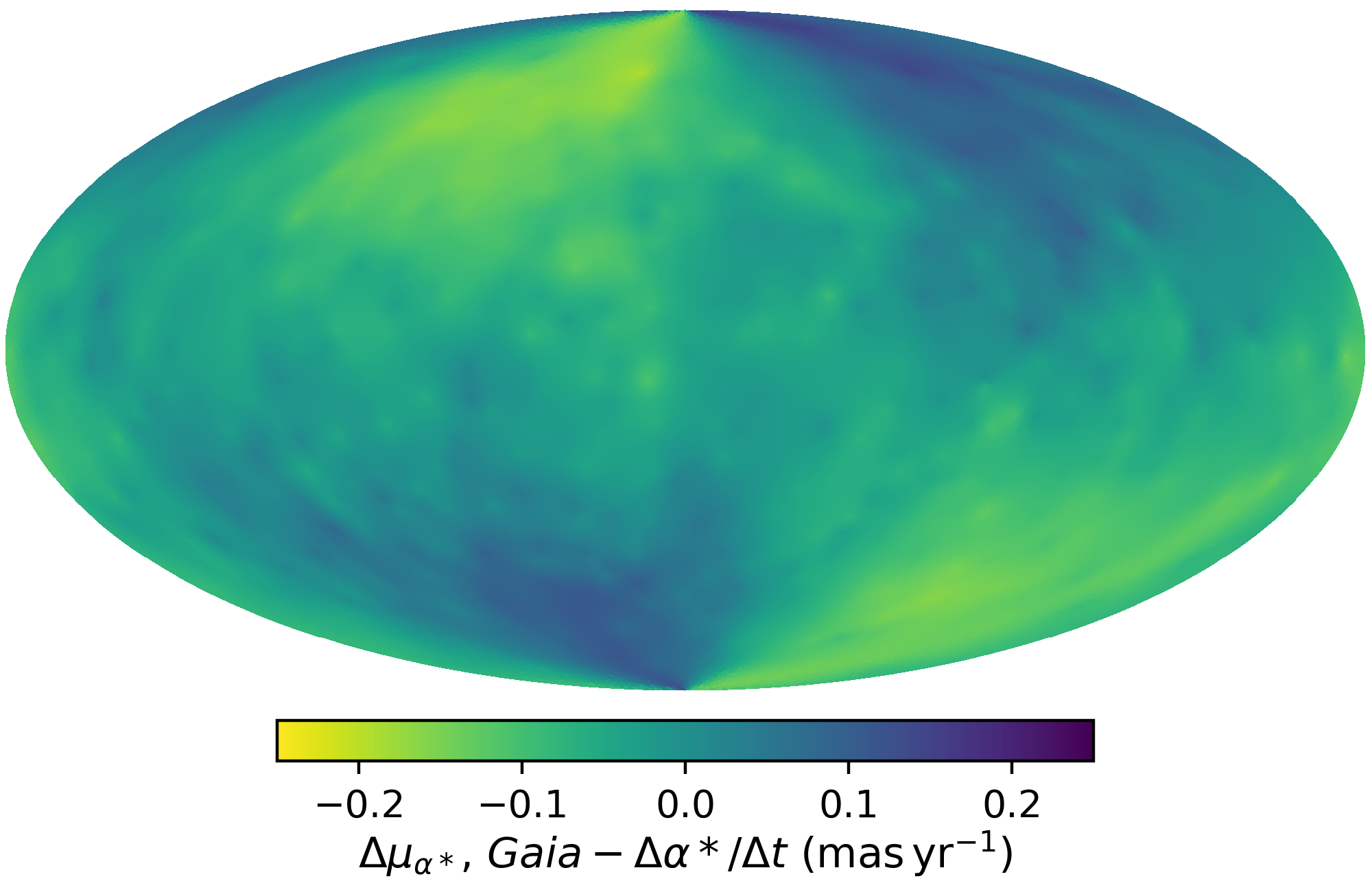}
    \includegraphics[width=0.5\linewidth]{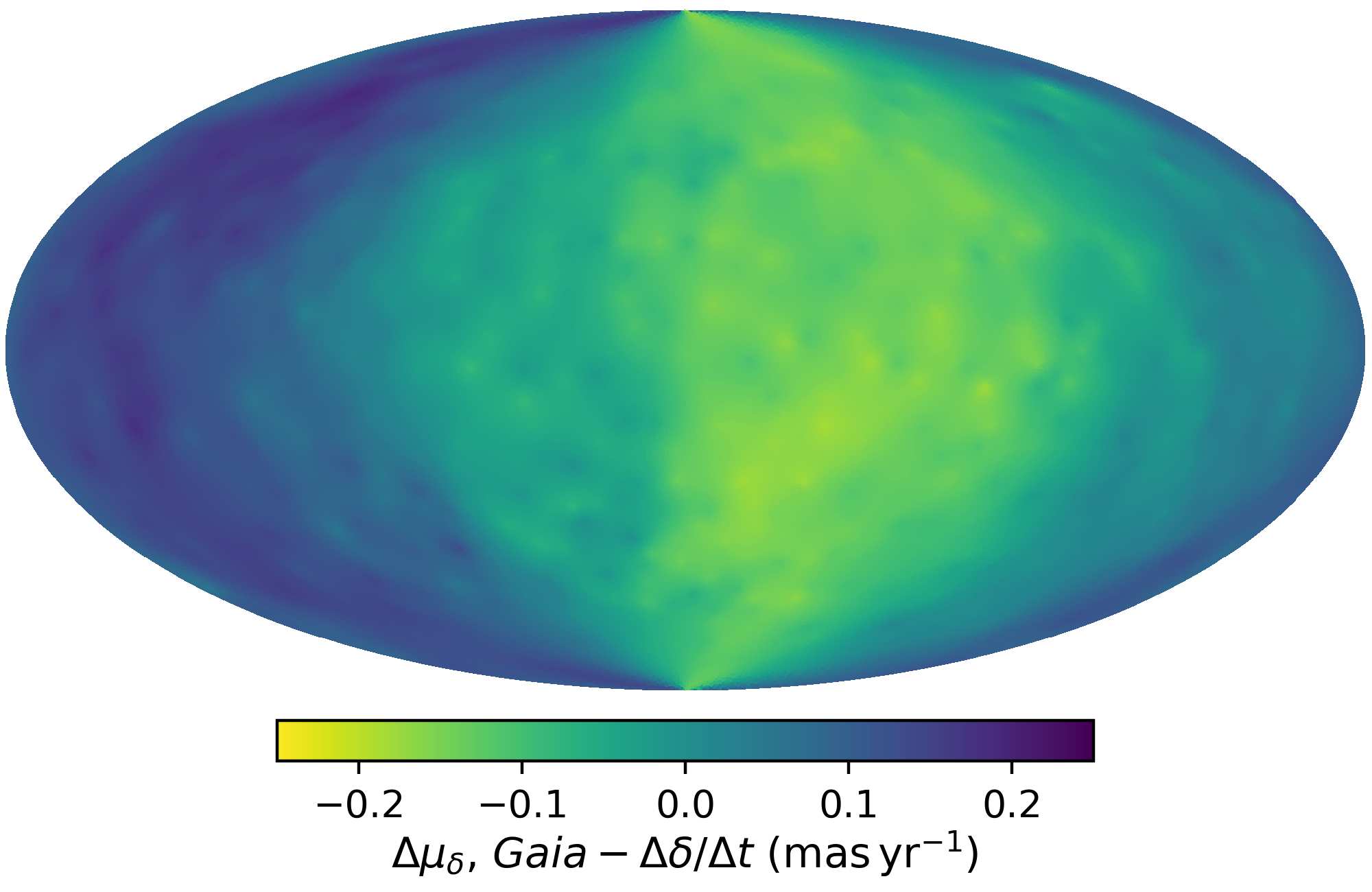}
    \vskip 0.1 truein
    \includegraphics[width=0.5\linewidth]{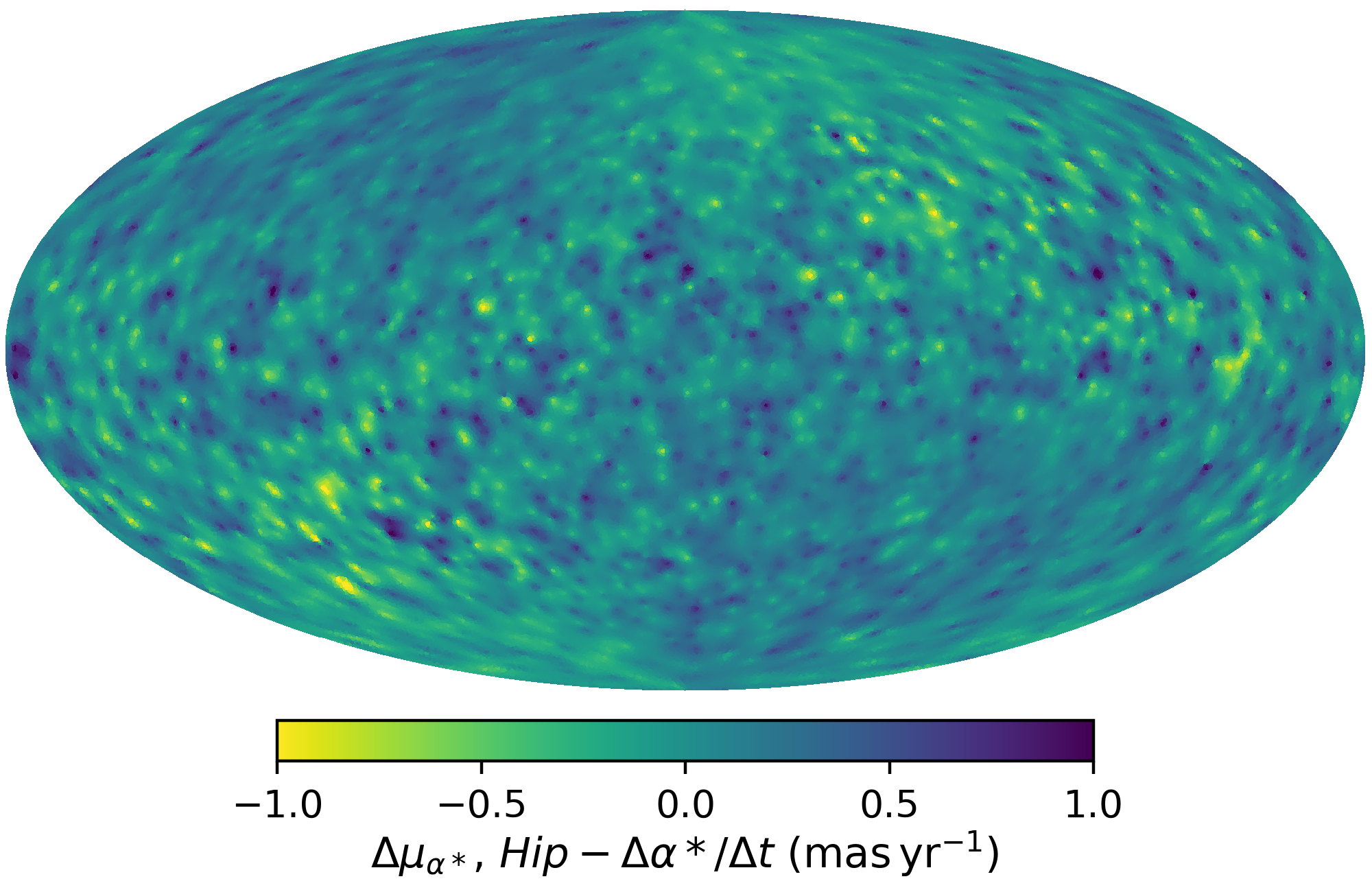}
    \includegraphics[width=0.5\linewidth]{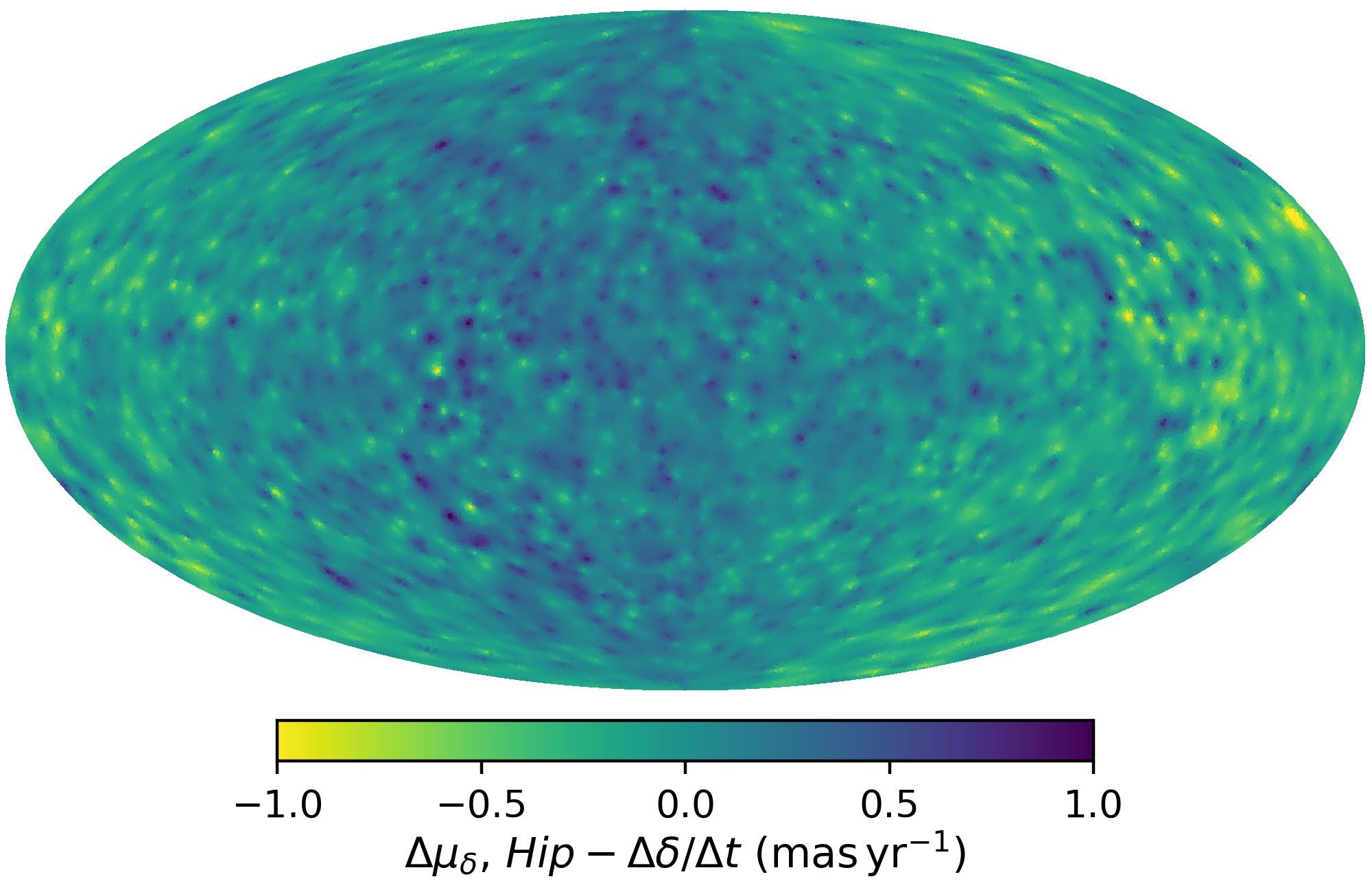}
    \caption{Local offsets between the {\it Gaia} DR2 proper motions and the scaled positional difference between {\it Hipparcos} and {\it Gaia} (top panels), and between {\it Hipparcos} and the scaled positional difference (bottom panels).  We obtained these offsets by separately fitting the proper motion residuals in each of the 920 regions shown in the top panel of Figure \ref{fig:map_coarse_tiled} omitting the stars with {\it Hipparcos} numbers ending in zero.  We then fit a Gaussian process regression using these stars as a cross-validation sample.  The maps give a higher log likelihood than a uniform rotation.  For the cross-validation set, the likelihood ratio of these maps to a uniform rotation is $e^{211}$ (for {\it Gaia}), and $e^{127}$ (for {\it Hipparcos}).  All maps are Hammer projections in equatorial coordinates: $(0,0)$ is at the center, east is left, and north is up.} 
    
    \label{fig:rotrates}
\end{figure*}

Figure \ref{fig:rotrates} shows the results projected into the local values of $\Delta \mu_{\alpha*}$ and $\Delta \mu_\delta$.  The differences between the {\it Gaia} proper motions and the scaled positional differences are dominated by an overall frame rotation (see Figure 4 of \citealt{Gaia_Astrometry_2018}).  The residuals, however, are highly significant: the likelihood ratio of the locally variable model to a model with uniform rotation is $e^{211}$ for the cross-validation sample.  The differences between the {\it Hipparcos} proper motions and the scaled positional differences show considerably more scatter, especially on smaller scales.  This scatter is even more highly significant than for {\it Gaia}, with a likelihood ratio of more than $e^{300}$ for our cross-validation sample.  

The likelihood ratio of the Gaussian process regression to uniform rotation for the full data set is just over $e^{3000}$ for the {\it Gaia} proper motions.  This is close to the tenth power of the cross-validation likelihood ratio ($e^{211}$).  Since the full sample is ten times the size of the cross-validation sample, this indicates that there is little overfitting.  The discrepancy between these numbers grows with a finer tiling of the sky.  The likelihood ratio for the full data set is around $e^{7000}$ for the {\it Hipparcos} proper motion differences, considerably more than the tenth power of the cross-validation sample's likelihood ratio of $e^{300}$.  Much of the likelihood improvement for the full sample in {\it Hipparcos} proper motion is probably due to overfitting.  However, small-scale systematics and correlated errors are still present at high significance.

We next move to the other parameters of the fit: the relative weights of the {\it Hipparcos} reductions and the error inflation terms.  We apply the offsets that we have fit in our Gaussian process regression before optimizing these other parameters in each region.  Again, we hold back 10\% of the stars as a cross-validation sample, and use these stars to optimize the parameters of our Gaussian process regression.

We find some evidence in favor of local variations in $f$ and $b$, the relative weights of the {\it Hipparcos} reductions and the {\it Hipparcos} error inflation term.  The likelihood ratios are around $e^{10}$ for the cross-validation sample after optimizing the model using the cross-validation stars.  This evidence is far weaker than for the frame rotation, and raises significant concerns with overfitting and underestimation of errors.  We therefore adopt a global fit of $f$ and $b$ to the cross-validation sample, obtaining values of $f=0.597$ and $b=0.199$.  

\begin{figure}
    \includegraphics[width=\linewidth]{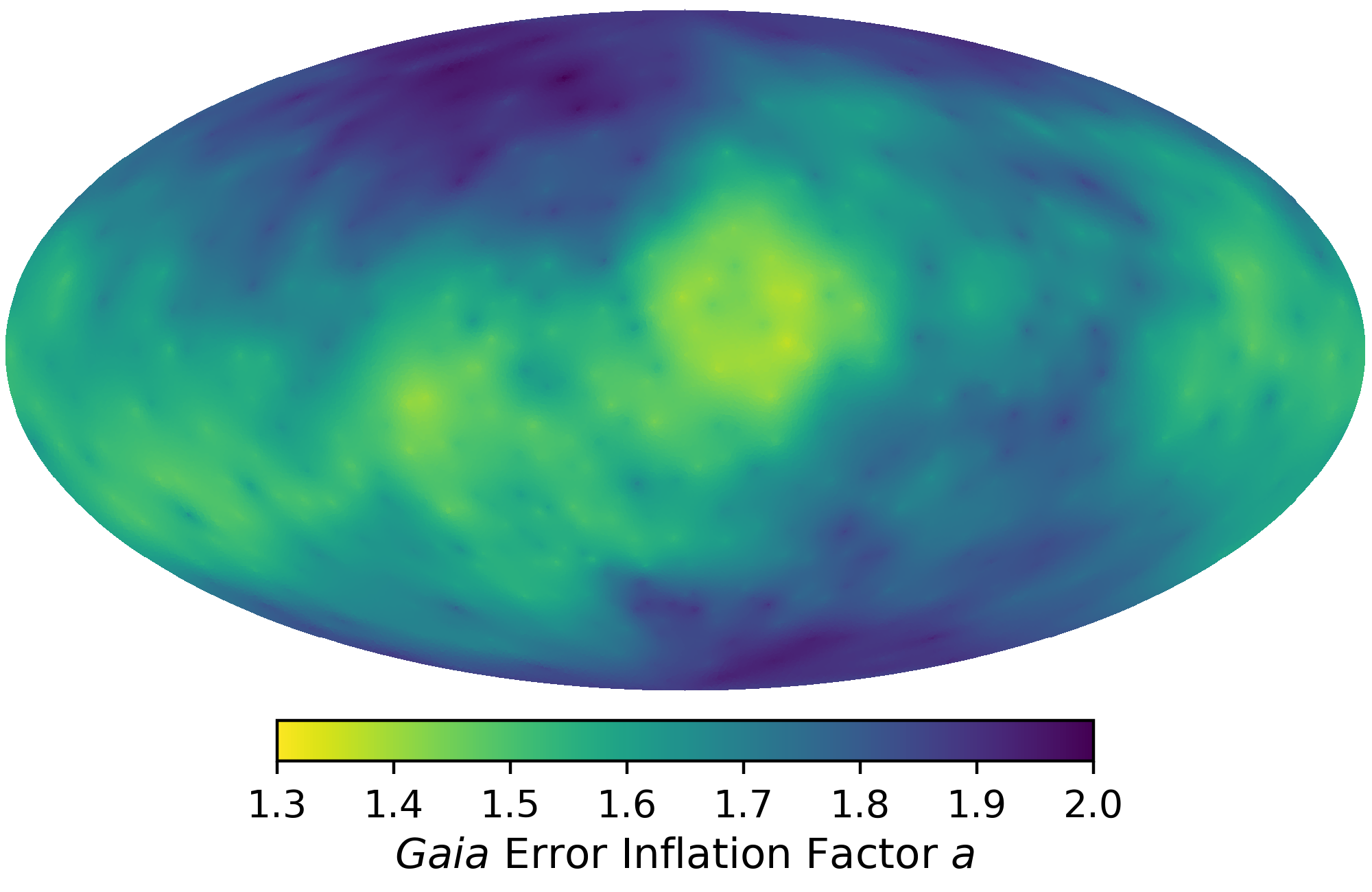}
    \caption{Map of the {\it Gaia} error inflation factor $a$ after fitting in 920 regions and using 10\% of stars as a cross-validation sample to optimize a Gaussian process regression model.  The likelihood ratio of this model to a constant $a$ is $e^{63}$ for the cross-validation sample, strong evidence in favor of spatial variation in the systematics of the {\it Gaia} uncertainties.}
    \label{fig:gaia_a_map}
\end{figure}

We can draw stronger conclusions for the {\it Gaia} error inflation factor $a$, shown in Figure \ref{fig:gaia_a_map}.  In this case, the relative likelihood of the spatially varying to the spatially uniform model is $e^{63}$ for the cross-validation sample, and $e^{782}$ for the full sample.  This is strong evidence that the ratio between the formal {\it Gaia} uncertainties and actual errors is spatially variable.  Our local fits for $a$ appear to suffer only mild overfitting.

The total improvement in likelihood of the optimized, locally variable model is nearly $e^{600}$ for our cross-validation data set, $\sim$35\,$\sigma$ evidence against uniform frame rotation rates, catalog weights, and error inflations.  
We use the maps shown in Figures \ref{fig:rotrates} 
and \ref{fig:gaia_a_map}, together with the fixed values $f=0.597$ and $b=0.199$, to construct the final catalog.  The evidence in their favor over the values in Table \ref{tab:parameterfits} is decisive.

\begin{figure*}
    \includegraphics[width=0.495\linewidth]{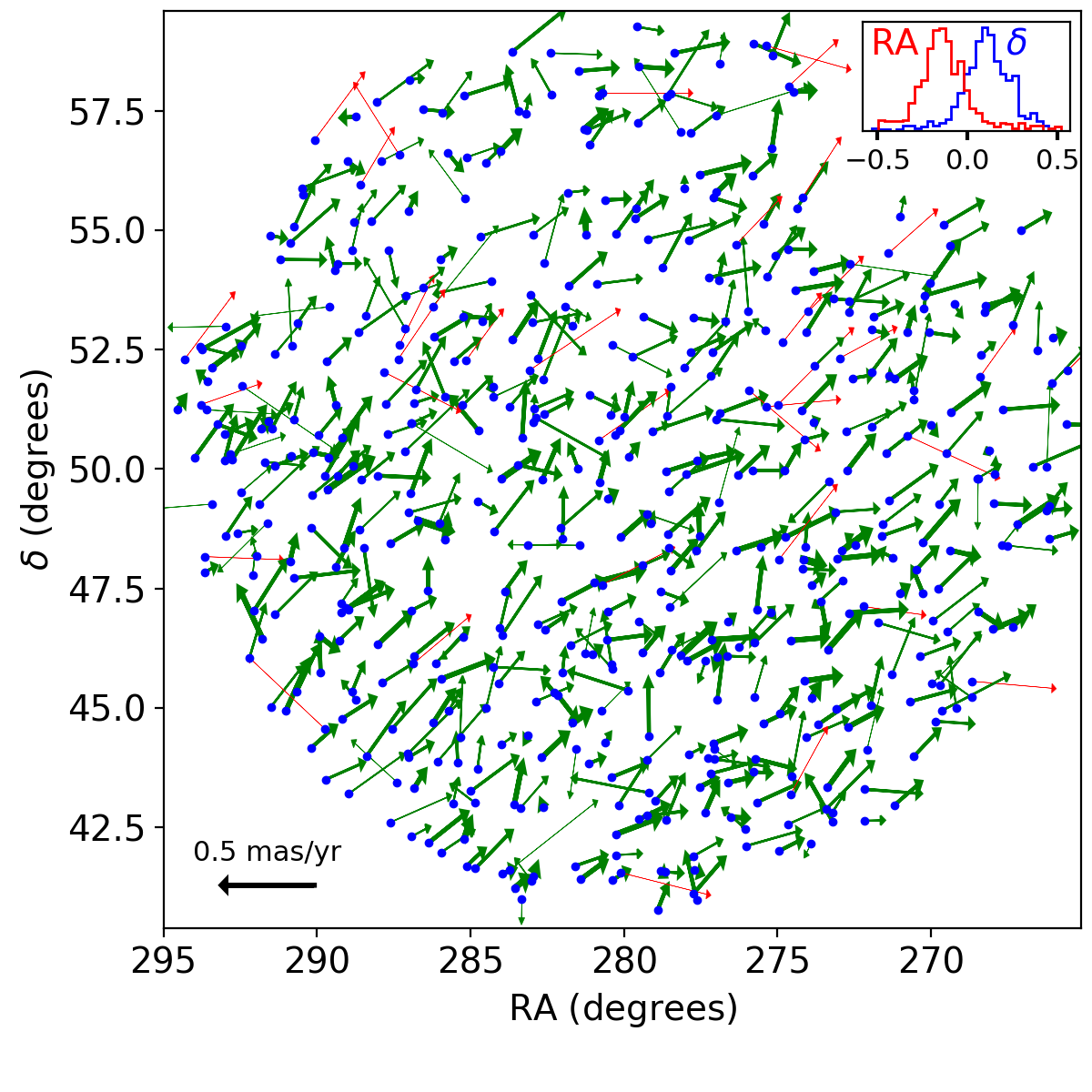}
    \includegraphics[width=0.495\linewidth]{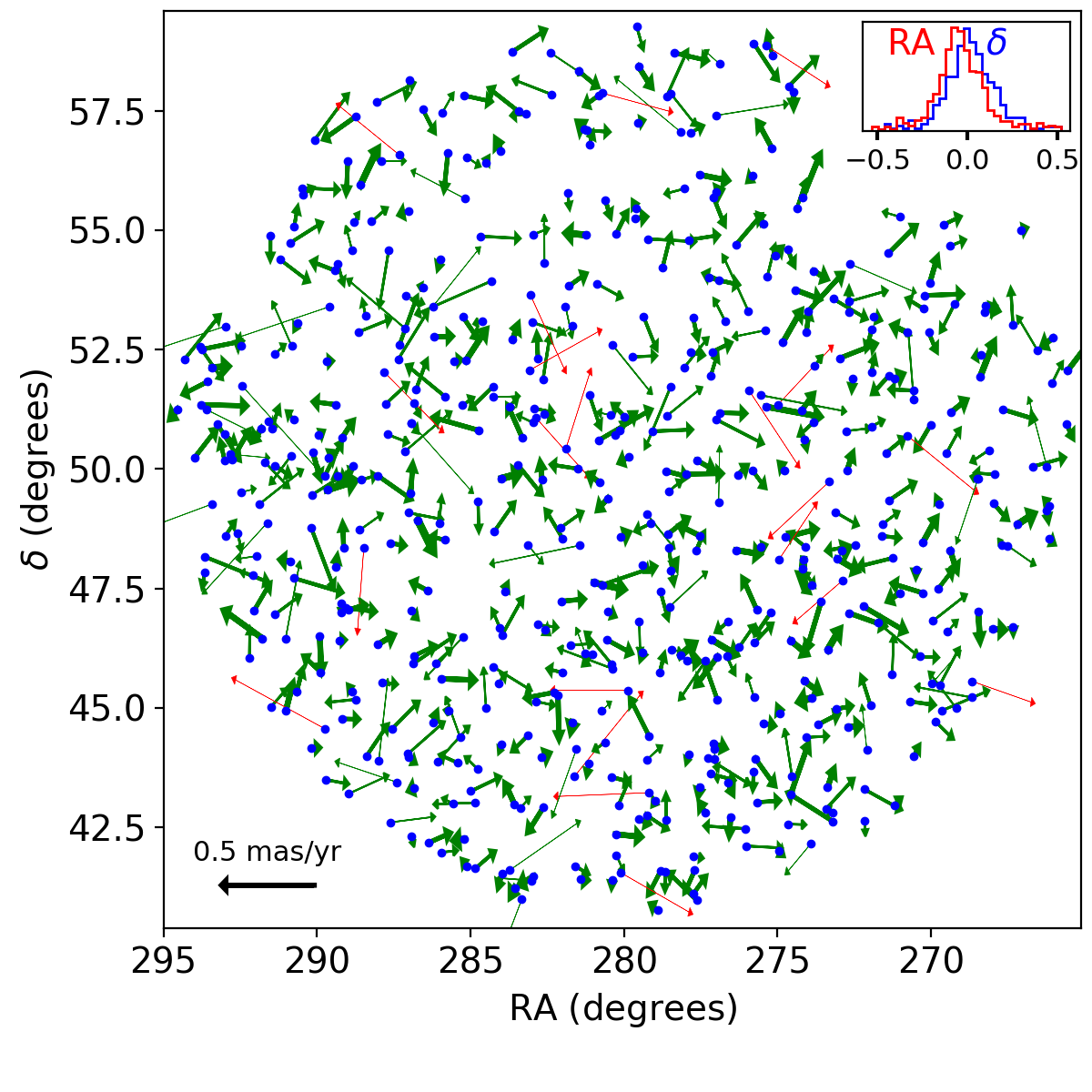}\\
    \centering\includegraphics[width=0.495\linewidth]{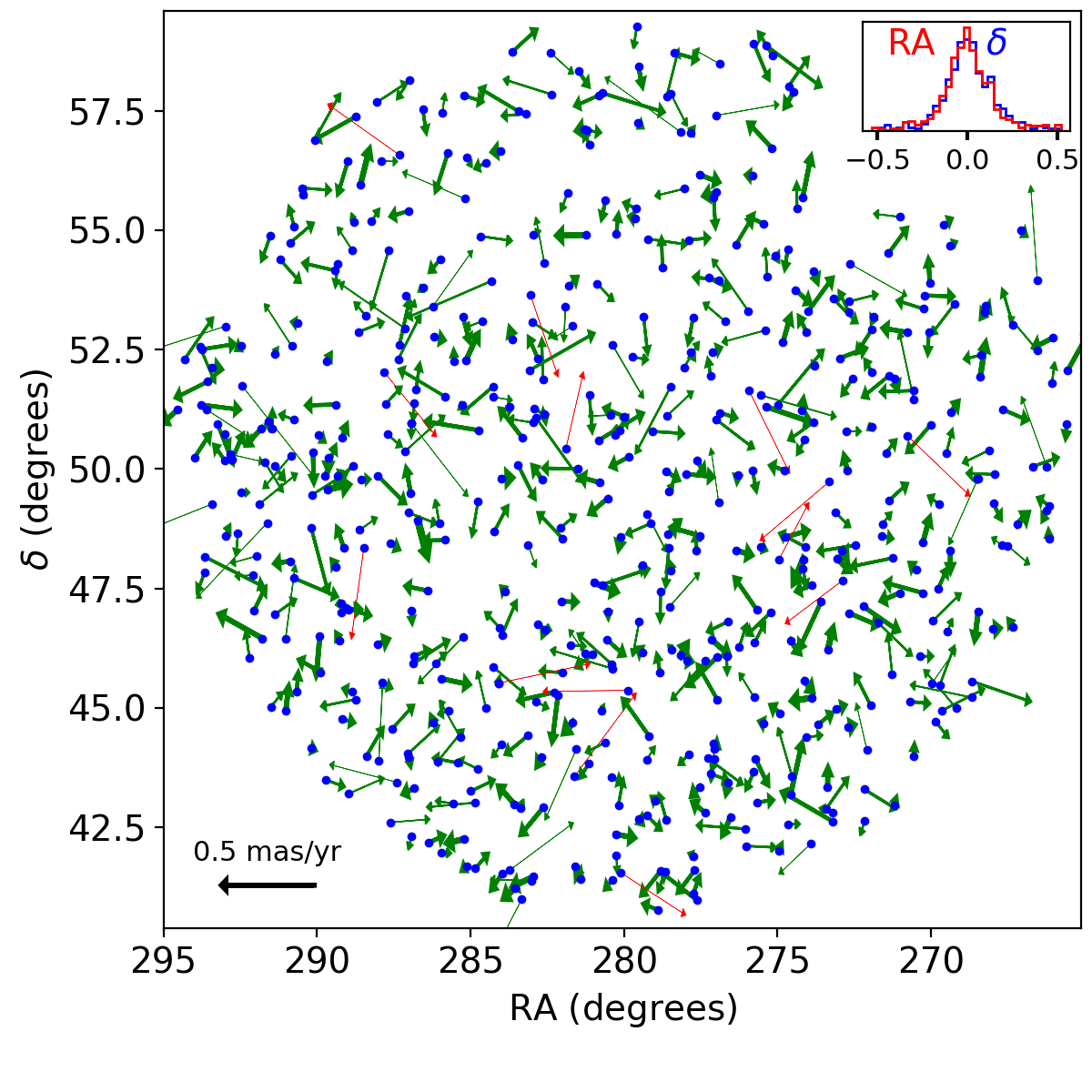}
    \caption{Residuals between the {\it Gaia} DR2 proper motions and the scaled {\it Hipparcos}--{\it Gaia} positional differences in a region of the sky with a particularly high offset between the catalogs' reference frames.  Only stars with residuals $\leq$0.5~mas\,yr$^{-1}$ in each direction are shown.  The top-left panel shows the results with no cross-calibration, the top-right panel shows the results using the global fit given in Table \ref{tab:parameterfits}, and the lower panel shows the results after applying the smooth, local fit derived in Section \ref{sec:localfit}.  The widths of the lines are proportional to their inverse variances.  Stars with more than an 80\% posterior probability of being outliers in our Gaussian mixture model (Equation \eqref{eq:gaussmix}) are shown with thin, red arrows regardless of their formal variances.  The residuals improve dramatically with the global calibration (which closely matches that calculated by the {\it Gaia} team; Figure 4 of \citealt{Gaia_Astrometry_2018}), but are better still with our smooth, local fit.  The residuals appear to be random across the field. }  
    \label{fig:demo_pmcorrect}
\end{figure*}

Figure \ref{fig:demo_pmcorrect} shows the differences between the {\it Gaia} DR2 proper motions and the {\it Hipparcos}--{\it Gaia} scaled positional differences in three cases: with no cross-calibration (top-left), with a global cross-calibration (top-right), and with the smooth local cross-calibration described here (bottom).  We show a region of the sky where the proper motion offset between the reference frames is particularly large.  The global calibration, nearly identical to the frame rotation rate found by \cite{Gaia_Astrometry_2018}, gives offsets that are qualitatively better than the raw catalogs.  Our smoothed, locally variable calibration is better still; this is particularly evident in the inset histograms.  The residuals in this case appear to be randomly oriented across the field.

\section{Final Construction and Structure of the Catalog} \label{sec:catalog}

Table \ref{tab:parameterfits} lists the best-fit parameters for a global cross-calibration of the {\it Hipparcos} and {\it Gaia} catalogs.  For our final catalog, we instead use the local calibration derived in the previous section and shown in Figures \ref{fig:rotrates} and \ref{fig:gaia_a_map}.  These local calibrations have a higher likelihood by a factor of almost $e^{600}$ in our cross-validation sample compared to the global calibration, or almost 35\,$\sigma$.  

We adopt the {\it Gaia} DR2 reference frame for our catalog; the published DR2 proper motions exactly match our values.  We do not calibrate to the scaled positional differences (our most precise measurement) because this would introduce a new reference frame into the literature, distinct from TGAS for many reasons including our use of a composite {\it Hipparcos} catalog.  \cite{Gaia_ICRF_2018} note that even for faint quasars, a reference frame that is strictly nonrotating for all subsets of the data is not yet possible.  Our scaled positional differences are similar to the values of the {\it Tycho}--{\it Gaia} astrometric solution \citep[TGAS,][]{TGAS_Astrometry_2016}, but with some important differences:
\begin{itemize}
    \item All of our values are locally calibrated to the DR2 reference frame; they are rotated relative to the TGAS frame (see Figure 4 of \cite{Gaia_Astrometry_2018}).  For the user who only measures differences in proper motion, the absolute reference frame is irrelevant.
    \item We adopt a composite {\it Hipparcos} catalog, use {\it Gaia} parallaxes to improve the other astrometric parameters, and use the {\it Hipparcos} positions at their central epochs.
    \item We adopt a slightly different model for error inflation.  
\end{itemize}
We also include the central epochs of the {\it Hipparcos} and {\it Gaia} position measurements (which we assume to also represent the central epochs of the proper motion measurements).  These epochs are close to 1991.25 and 2015.5, but typically differ from these values by $\sim$0.1-0.2 years.

\begin{figure}
    \centering
    \includegraphics[width=\linewidth]{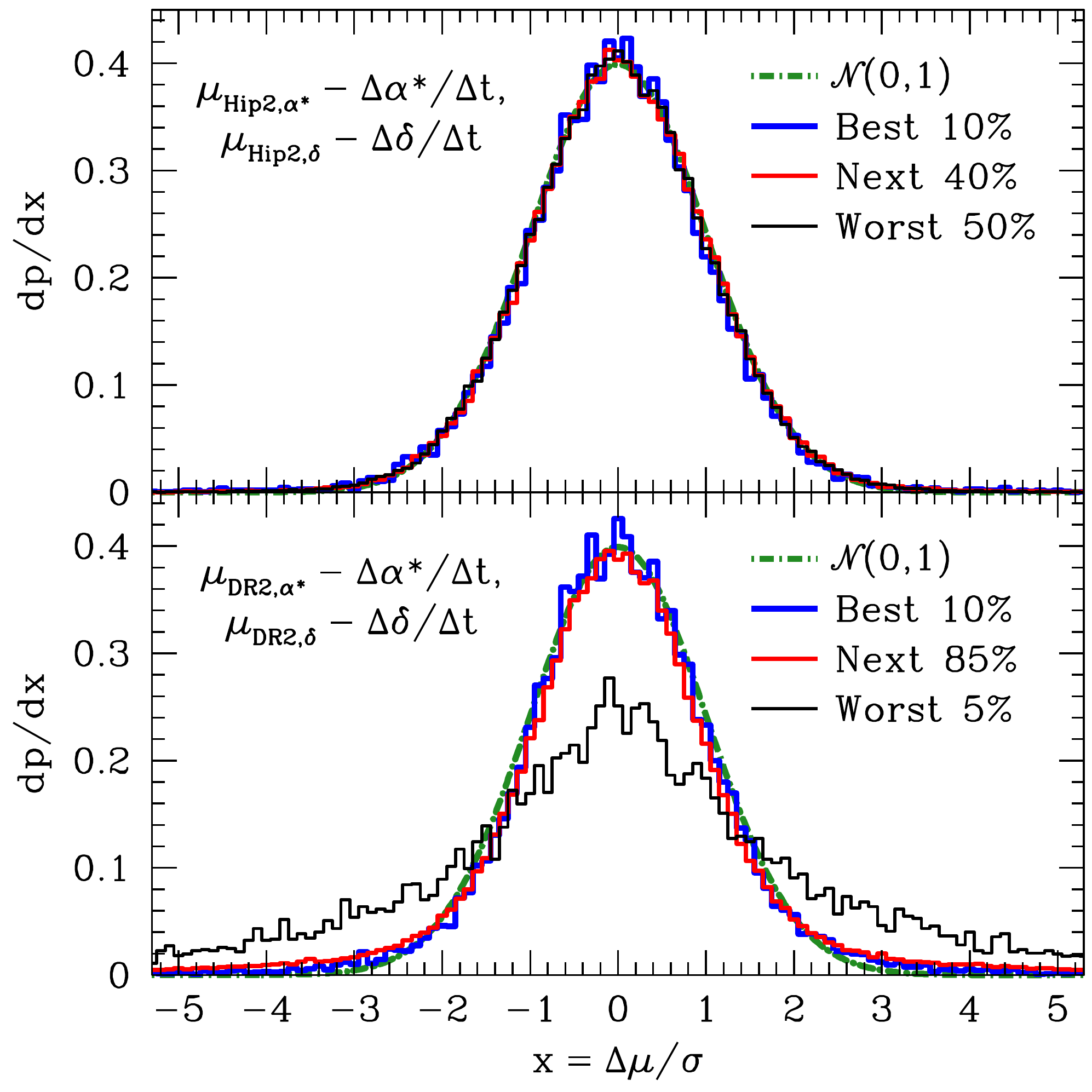}
    \caption{Normalized residuals of the differences in proper motion, both in right ascension and in declination.  The stars shown here are the same $\sim$93,000 as in Figure \ref{fig:baseline_uncertainties}, but in this case we have applied the full cross-calibration described in Section \ref{sec:localfit}.  The distributions now look accurately Gaussian with the correct variance.  The lowest-precision {\it Gaia} stars show heavy tails, indicating that their uncertainties remain underestimated by as much as a factor of $\sim$2.  These stars have calibrated {\it Gaia} DR2 proper motion uncertainties $\gtrsim$0.7~mas\,yr$^{-1}$; many of them are exceptionally bright.  With forthcoming improvements in the treatment of saturated stars and a full correction of the DOF bug \citep{Gaia_Astrometry_2018}, future {\it Gaia} data releases will be even better.}
    \label{fig:final_uncertainties}
\end{figure}

Our {\it Hipparcos} proper motions differ from previously published values due to our composite catalog, our local frame rotation, and our incorporation of {\it Gaia} DR2 parallaxes.  Figures \ref{fig:hip_err} and \ref{fig:final_uncertainties} show that these proper motions have Gaussian residuals with the expected uncertainties.  
Our {\it Hipparcos} astrometry also combines the two reductions of the raw data in an optimal way based both on more than 90,000 stars showing negligible acceleration, and on a smaller sample of 9,000 stars for cross-validation of a locally variable solution.  We have then placed our {\it Hipparcos} astrometry in the DR2 reference frame to enable direct comparisons between {\it Hipparcos} and {\it Gaia} measurements.

With the cross-calibration complete, we return to our initial cross-matched catalog, described in Section \ref{sec:xmatch}.  This catalog has 122,666 potential matches in DR2 to 116,074 stars in {\it Hipparcos}.  We take the best DR2 match to each {\it Hipparcos} star, i.e, the one with the lowest $\chi^2$ residual in the proper motions after calibrating coordinates and inflating uncertainties.  This yields one DR2 star for each {\it Hipparcos} star, with a potential match for 98\% of the {\it Hipparcos} catalog.  A few hundred of these matches are spurious.  We reject any potential matches with position changes corresponding to proper motions above 11$^{\prime\prime}\,{\rm yr}^{-1}$ (i.e.~anything faster than Barnard's Star).  We also reject any potential matches with $\chi^2>1.9 \times 10^5$.  This cut is ad-hoc, but above it, every potential match appears to be spurious.  Near the cut the catalog contains a mix of spurious and real matches, while at a factor of three below the cut (corresponding to about 15 stars below the cut) most of the matches are real.  This trimmed catalog contains 115,662 entries; we estimate that $\sim$100 are spurious.  We do not clean the catalog by hand beyond this level, but leave the final vetting to the user.  A few stars present in the catalog also have problematic astrometry.  
HIP 10529, for example, has a parallax of $\sim$50\,mas and proper motions of several hundred mas\,yr$^{-1}$ in both {\it Hipparcos} reductions, but a parallax of 4\,mas and a proper motion below 10\,mas\,yr$^{-1}$ in DR2 (the {\it Hipparcos} reductions appear to be corrupted by the nearby HIP 10531).  

Figure \ref{fig:final_uncertainties} shows our final distribution of normalized proper motion differences for nearly 93,000 stars consistent at 10\,$\sigma$ with zero astrometric acceleration, after applying the full correction described in Section \ref{sec:localfit}.  Figure \ref{fig:baseline_uncertainties} uses the same stars but assumes $\omega=0$, $b=0$, $a=1$, and $f=1$.  All distributions, with the exception of the lowest-precision {\it Gaia} stars (many of which are exceptionally bright, and actually have better proper motions from {\it Hipparcos} than from {\it Gaia} DR2), now look accurately Gaussian with the correct variance.  The 5\% threshold separating stars belonging to the red and black {\it Gaia} histograms corresponds to a calibrated proper motion uncertainty of $\sim$0.7~mas\,yr$^{-1}$.  The heavy tails in the very low-precision {\it Gaia} stars indicate that their errors may be underestimated by as much as a factor of $\sim$2.  This may be partially due to {\it Gaia}'s treatment of saturated stars, which will improve in future data releases.  The nature of the correction applied for the DOF bug also means that uncertainties of individual stars may be incorrect \citep{Gaia_Astrometry_2018}; this will be fixed in the next data release. 
With these caveats, Figure \ref{fig:final_uncertainties} demonstrates the overall reliability of the calibrated errors, and thus the suitability of our catalog for fitting accelerations and orbits.  

In addition to the proper motions and epochs, we report the covariance matrices for each set of proper motions.  The covariance between any two of the three sets of proper motions is very nearly zero:  our propagation of positional coordinates to their central epochs renders the covariance between position and proper motion exactly zero in a given coordinate (either right ascension or declination).  

For very nearby stars, we also address the fact that uniform motion in space does not produce uniform motion in celestial coordinates.  We use the {\it Gaia} DR2 radial velocities (where available) together with the DR2 positions and proper motions to predict the departure from uniform motion on the sphere at the appropriate {\it Hipparcos} epoch (this is distinct from the change in parallax mentioned in Section \ref{sec:gaia_plx}).  We then correct the {\it Hipparcos} proper motion and the positional difference for this effect.  It is negligible for the vast majority of stars.  

Tables \ref{tab:hgca_1}, \ref{tab:hgca_2}, and \ref{tab:hgca_3} show the structure of our catalog.  The full catalog is available electronically; these tables show only the first ten rows.  The tabulated proper motions $\mu_{\alpha*}$ and $\mu_\delta$ are related to the published catalog values as follows.  Using $\mu_{\alpha*,{\rm vL}}$ to represent the value of proper motion in right ascension in the \cite{vanLeeuwen_2007} catalog, $\mu_{\alpha*,{\rm ESA}}$ to represent the same value in the \cite{ESA_1997} catalog and $f=0.597$ to be the relative weightings,
\begin{equation}
    \mu_{\alpha*,H} = f \mu_{\alpha*,{\rm vL}} + (1 - f)\mu_{\alpha*,{\rm ESA}} + \xi_{\alpha*,H} + 2 \gamma_{\alpha*}.
    \label{eq:crosscal}
\end{equation}
Here, $\xi_{\alpha*,H}$ is the local cross-calibration of the catalogs derived from the data shown in Figure \ref{fig:rotrates} and $\gamma_{\alpha*}$ is a first-order correction for nonlinear proper motion of a star moving uniformly through space.  A similar equation applies for declination.  The term $\gamma_{\alpha*}$ is multiplied by two so that $\gamma_{\alpha*}$ itself is the difference between the mean proper motion (centered on the {\it Hipparcos} and {\it Gaia} epochs) and the instantaneous proper motion at either epoch.  For the scaled positional difference, we have the relation
\begin{equation}
    \mu_{\alpha*,HG} = \frac{f \alpha_{\rm vL} + (1 - f)\alpha_{\rm ESA} - \alpha_G}{t_G - (f t_{\rm vL} + (1 - f)t_{\rm ESA})}\cos \delta + \xi_{\alpha*,HG} + \gamma_{\alpha*}
    \label{eq:crosscal_hg}
\end{equation}
where $\alpha$ is the measured right ascension, $\delta$ is the measured declination (we take the average of DR2 and {\it Hipparcos}), $t_{\rm vL}$ is the central epoch of the proper motion of the \cite{vanLeeuwen_2007} Hipparcos reduction, and $\xi_{\alpha*,HG}$ is another local cross-calibration of the catalogs derived from the data shown in Figure \ref{fig:rotrates}.  The published proper motions include all of our derived corrections.

Three of the first ten stars, HIP~1, HIP~2, and HIP~7, show highly significant accelerations.  HIP~2 was reduced by {\it Hipparcos} as a multiple star with an orbital solution \citep{ESA_1997,vanLeeuwen_2007}; it also shows highly significant excess astrometric noise in {\it Gaia} DR2.  Neither HIP~1 nor HIP~7 was treated as a multiple in the {\it Hipparcos} reductions.  Neither shows excess astrometric noise in {\it Gaia} DR2, nor are they listed as multiple stars in Simbad, nor do they have close neighbors in DR2 with similar parallax.  Stars like these are excellent candidates for follow-up to discover and measure the masses of faint companions.

\begin{deluxetable*}{lcccccccccccr}
\tablewidth{0pt}
\tablecaption{The {\it Hipparcos}--{\it Gaia} Catalog of Accelerations: {\it Hipparcos} Proper Motions}
\tablehead{
    \colhead{{\it Hipparcos}} &
    \colhead{{\it Gaia} DR2} &
    \colhead{$\mu_{\alpha*,H}$\tablenotemark{$\dagger$}} &
    \colhead{$\sigma_{\alpha*,H}$} &
    \colhead{$\mu_{\delta,H}$\tablenotemark{$\dagger$}} &
    \colhead{$\sigma_{\delta,H}$} &
    \colhead{Corr} &
    \colhead{$t_{\alpha*,H}$} &
    \colhead{$t_{\delta,H}$} &
    \colhead{$\xi_{\alpha*,H}$} &
    \colhead{$\xi_\delta,H$} &
    \colhead{$2\gamma_{\alpha*}$} &
    \colhead{$2\gamma_{\delta}$} 
    \\
    \colhead{Number} &
    \colhead{Source ID} &
    \multicolumn{2}{c}{mas\,yr$^{-1}$} &
    \multicolumn{2}{c}{mas\,yr$^{-1}$} &
    \colhead{} &
    \multicolumn{2}{c}{year} &
    \multicolumn{2}{c}{mas\,yr$^{-1}$} &
    \multicolumn{2}{c}{mas\,yr$^{-1}$}
    }
\startdata
  1 & 2738327528519591936 & $  -5.02$ & $1.31$ & $  -1.20$ & $0.80$ & $ 0.35$& 1991.55 & 1991.28 & $-0.17$ & $ 0.28$ & $0.00$ & $0.00$ \\
  2 & 2341871673090078592 & $ 183.12$ & $1.41$ & $  -0.89$ & $0.78$ & $ 0.14$& 1991.47 & 1991.42 & $-0.22$ & $ 0.37$ & $0.00$ & $0.00$ \\
  3 & 2881742980523997824 & $   4.57$ & $0.49$ & $  -2.89$ & $0.41$ & $ 0.18$& 1990.85 & 1991.05 & $-0.07$ & $ 0.32$ & $0.00$ & $0.00$ \\
  4 & 4973386040722654336 & $  62.69$ & $0.60$ & $   0.55$ & $0.57$ & $-0.16$& 1991.01 & 1991.18 & $-0.02$ & $ 0.32$ & $0.00$ & $0.00$ \\
  5 & 2305974989264598272 & $   1.84$ & $0.63$ & $   8.64$ & $0.70$ & $ 0.09$& 1991.10 & 1991.48 & $-0.18$ & $ 0.07$ & $0.00$ & $0.00$ \\
  6 & 2740326852975975040 & $ 223.93$ & $5.77$ & $ -14.03$ & $3.18$ & $ 0.25$& 1991.34 & 1991.26 & $-0.09$ & $ 0.34$ & $0.00$ & $0.00$ \\
  7 & 2846308881856186240 & $-208.11$ & $1.09$ & $-200.78$ & $0.78$ & $ 0.41$& 1991.29 & 1991.23 & $-0.12$ & $ 0.42$ & $0.00$ & $0.00$ \\
  8 & 2853169937491828608 & $  19.07$ & $1.32$ & $  -6.11$ & $0.77$ & $ 0.05$& 1991.58 & 1991.46 & $-0.07$ & $ 0.41$ & $0.00$ & $0.00$ \\
  9 & 2880160886370458368 & $  -6.85$ & $1.05$ & $   8.75$ & $0.64$ & $ 0.11$& 1991.26 & 1991.20 & $-0.19$ & $ 0.28$ & $0.00$ & $0.00$ \\
 10 & 4976500987226833024 & $  41.93$ & $0.96$ & $  40.82$ & $0.80$ & $-0.11$& 1991.24 & 1991.41 & $-0.04$ & $ 0.30$ & $0.00$ & $0.00$ 
 
 \enddata
 \tablenotetext{$\dagger$}{Values include all local and nonlinearity corrections (e.g.~$\xi$ and $\gamma$, see Equation \eqref{eq:crosscal}).}
 \label{tab:hgca_1}
\end{deluxetable*}

\begin{deluxetable*}{lcccccccccr}
\tablewidth{0pt}
\tablecaption{The {\it Hipparcos}--{\it Gaia} Catalog of Accelerations: {\it Gaia} DR2--{\it Hipparcos} Scaled Position Differences}
\tablehead{
    \colhead{{\it Hipparcos}} &
    \colhead{{\it Gaia} DR2} &
    \colhead{$\mu_{\alpha*,HG}$\tablenotemark{$\dagger$}} &
    \colhead{$\sigma_{\alpha*,HG}$} &
    \colhead{$\mu_{\delta,HG}$\tablenotemark{$\dagger$}} &
    \colhead{$\sigma_{\delta,HG}$} &
    \colhead{Corr} &
    \colhead{$\xi_{\alpha*,HG}$} &
    \colhead{$\xi_{\delta,HG}$} &
    \colhead{$\gamma_{\alpha*}$} &
    \colhead{$\gamma_{\delta}$} 
    \\
    \colhead{Number} &
    \colhead{Source ID} &
    \multicolumn{2}{c}{mas\,yr$^{-1}$} &
    \multicolumn{2}{c}{mas\,yr$^{-1}$} &
    \colhead{} &
    \multicolumn{2}{c}{mas\,yr$^{-1}$} &
    \multicolumn{2}{c}{mas\,yr$^{-1}$} 
    }
\startdata
  1 & 2738327528519591936 & $  -6.006$ & $0.055$ & $  -4.988$ & $0.030$ & $ 0.34$& $-0.108$ & $ 0.097$ & $0.000$ & $0.000$ \\
  2 & 2341871673090078592 & $ 181.367$ & $0.049$ & $  -0.316$ & $0.028$ & $ 0.13$& $-0.106$ & $ 0.113$ & $0.000$ & $0.001$ \\
  3 & 2881742980523997824 & $   5.811$ & $0.019$ & $  -2.365$ & $0.015$ & $ 0.07$& $ 0.046$ & $ 0.102$ & $0.000$ & $0.000$ \\
  4 & 4973386040722654336 & $  61.911$ & $0.020$ & $   1.409$ & $0.023$ & $-0.27$& $-0.078$ & $ 0.111$ & $0.000$ & $0.000$ \\
  5 & 2305974989264598272 & $   0.915$ & $0.024$ & $   8.844$ & $0.024$ & $ 0.05$& $-0.077$ & $ 0.081$ & $0.000$ & $0.000$ \\
  6 & 2740326852975975040 & $ 223.069$ & $0.179$ & $ -11.472$ & $0.097$ & $ 0.37$& $-0.096$ & $ 0.096$ & $0.000$ & $0.000$ \\
  7 & 2846308881856186240 & $-211.152$ & $0.041$ & $-196.918$ & $0.032$ & $ 0.31$& $-0.031$ & $ 0.098$ & $0.002$ & $0.000$ \\
  8 & 2853169937491828608 & $  18.707$ & $0.053$ & $  -6.494$ & $0.034$ & $ 0.03$& $ 0.001$ & $ 0.113$ & $0.000$ & $0.000$ \\
  9 & 2880160886370458368 & $  -6.002$ & $0.036$ & $   9.358$ & $0.023$ & $ 0.01$& $ 0.038$ & $ 0.100$ & $0.000$ & $0.000$ \\
 10 & 4976500987226833024 & $  42.260$ & $0.029$ & $  40.938$ & $0.030$ & $-0.09$& $-0.078$ & $ 0.111$ & $0.000$ & $0.000$ 
 \enddata
 \tablenotetext{$\dagger$}{Values include all local and nonlinearity corrections (e.g.~$\xi$ and $\gamma$, see Equation \eqref{eq:crosscal_hg}).}
  \label{tab:hgca_2}
\end{deluxetable*}

\begin{deluxetable*}{lccccccccr}
\tablewidth{0pt}
\tablecaption{The {\it Hipparcos}--{\it Gaia} Catalog of Accelerations: {\it Gaia} DR2 Proper Motions}
\tablehead{
    \colhead{{\it Gaia} DR2} &
    \colhead{$\mu_{\alpha*,G}$\tablenotemark{$\dagger$}} &
    \colhead{$\sigma_{\alpha*,G}$} &
    \colhead{$\mu_{\delta,G}$\tablenotemark{$\dagger$}} &
    \colhead{$\sigma_{\delta,G}$} &
    \colhead{Corr\tablenotemark{$\dagger$}} &
    \colhead{$t_{\alpha*,G}$} &
    \colhead{$t_{\delta,G}$} 
    \\
    \colhead{Source ID} &
    \multicolumn{2}{c}{mas\,yr$^{-1}$} &
    \multicolumn{2}{c}{mas\,yr$^{-1}$} &
    \colhead{} &
    \multicolumn{2}{c}{year}
    }
\startdata
2738327528519591936 & $  -0.656$ & $0.176$ & $  -5.132$ & $0.071$ & $ 0.17$ & 2015.60 & 2015.37 \\
2341871673090078592 & $ 163.520$ & $0.798$ & $  -2.362$ & $0.609$ & $ 0.11$ & 2015.37 & 2015.38 \\
2881742980523997824 & $   5.940$ & $0.141$ & $  -2.299$ & $0.100$ & $-0.16$ & 2015.75 & 2015.65 \\
4973386040722654336 & $  61.800$ & $0.072$ & $   1.438$ & $0.077$ & $-0.21$ & 2015.60 & 2015.58 \\
2305974989264598272 & $   0.967$ & $0.097$ & $   8.782$ & $0.127$ & $ 0.17$ & 2015.96 & 2015.34 \\
2740326852975975040 & $ 223.124$ & $0.123$ & $ -11.367$ & $0.064$ & $ 0.26$ & 2015.45 & 2015.24 \\
2846308881856186240 & $-206.469$ & $0.136$ & $-195.796$ & $0.054$ & $-0.14$ & 2015.37 & 2015.21 \\
2853169937491828608 & $  18.526$ & $0.279$ & $  -6.549$ & $0.128$ & $ 0.08$ & 2015.83 & 2015.53 \\
2880160886370458368 & $  -5.973$ & $0.114$ & $   9.447$ & $0.066$ & $-0.23$ & 2015.60 & 2015.67 \\
4976500987226833024 & $  42.239$ & $0.074$ & $  41.050$ & $0.078$ & $-0.28$ & 2015.59 & 2015.62 
\enddata
\tablenotetext{$\dagger$}{Values are identical to those published in {\it Gaia} DR2 \citep{Gaia_Astrometry_2018}.}
 \label{tab:hgca_3}
\end{deluxetable*}

\begin{deluxetable*}{lcr}
\tablewidth{0pt}
\tablecaption{The {\it Hipparcos}--{\it Gaia} Catalog of Accelerations: Description of Catalog Contents}
\tablehead{
    \colhead{Parameter Name} &
    \colhead{Units} &
    \colhead{Description}
    }
\startdata
${\tt hip\_id}$ & & {\it Hipparcos} identification number \\
${\tt gaia\_source\_id}$ & & {\it Gaia} DR2 source identification number \\
${\tt gaia\_ra}$ & degrees & {\it Gaia} DR2 measured right ascension \\
${\tt gaia\_dec}$ & degrees & {\it Gaia} DR2 measured declination \\
${\tt radial\_velocity}$ & km\,s$^{-1}$ & {\it Gaia} DR2 measured radial velocity \\
${\tt radial\_velocity\_error}$ & km\,s$^{-1}$ & {\it Gaia} DR2 radial velocity uncalibrated standard error \\
${\tt gaia\_parallax}$ & mas & {\it Gaia} DR2 parallax \\
${\tt gaia\_parallax\_error}$ & mas & {\it Gaia} DR2 parallax standard error \\
${\tt pmra\_gaia}$ & mas\,yr$^{-1}$ & {\it Gaia} DR2 proper motion in right ascension, $d\alpha/dt \cos \delta$ \\
${\tt pmra\_gaia\_error}$ & mas\,yr$^{-1}$ & Calibrated uncertainty in ${\tt pmra\_gaia}$ \\
${\tt pmdec\_gaia}$ & mas\,yr$^{-1}$ & {\it Gaia} DR2 proper motion in declination \\
${\tt pmdec\_gaia\_error}$ & mas\,yr$^{-1}$ & Calibrated uncertainty in ${\tt pmdec\_gaia}$ \\
${\tt pmra\_pmdec\_gaia}$ & & Correlation between ${\tt pmra\_gaia}$ and ${\tt pmdec\_gaia}$ \\
${\tt pmra\_hg}$ & mas\,yr$^{-1}$ & Calibrated proper motion in right ascension from the {\it Hipparcos}--{\it Gaia} positional difference \\
${\tt pmra\_hg\_error}$ & mas\,yr$^{-1}$ & Calibrated uncertainty in ${\tt pmra\_hg}$ \\
${\tt pmdec\_hg}$ & mas\,yr$^{-1}$ & Calibrated proper motion in declination from the {\it Hipparcos}--{\it Gaia} positional difference \\
${\tt pmdec\_hg\_error}$ & mas\,yr$^{-1}$ & Calibrated uncertainty in  ${\tt pmdec\_hg}$ \\
${\tt pmra\_pmdec\_hg}$ & & Correlation between ${\tt pmra\_hg}$ and ${\tt pmdec\_hg}$ \\
${\tt pmra\_hip}$ & mas\,yr$^{-1}$ & Calibrated proper motion in right ascension from the composite {\it Hipparcos} catalog \\
${\tt pmra\_hip\_error}$ & mas\,yr$^{-1}$ & Calibrated uncertainty in ${\tt pmra\_hip}$ \\
${\tt pmdec\_hip}$ & mas\,yr$^{-1}$ & Calibrated proper motion in declination from the composite {\it Hipparcos} catalog \\
${\tt pmdec\_hip\_error}$ & mas\,yr$^{-1}$ & Calibrated uncertainty in  ${\tt pmdec\_hip}$ \\
${\tt pmra\_pmdec\_hip}$ & & Correlation between ${\tt pmra\_hip}$ and ${\tt pmdec\_hip}$ \\
${\tt epoch\_ra\_gaia}$ & year & Central epoch of {\it Gaia} DR2 right ascension measurement \\
${\tt epoch\_dec\_gaia}$ & year & Central epoch of {\it Gaia} DR2 declination measurement \\
${\tt epoch\_ra\_hip}$ & year & Central epoch of {\it Hipparcos} right ascension measurement \\
${\tt epoch\_dec\_hip}$ & year & Central epoch of {\it Hipparcos} declination measurement \\
${\tt crosscal\_pmra\_hg}$ & mas\,yr$^{-1}$ & Difference in ${\tt pmra\_hg}$ from the catalog-computed value: $\xi_{\alpha*,HG}$ from Table \ref{tab:hgca_2} \\
${\tt crosscal\_pmdec\_hg}$ & mas\,yr$^{-1}$ & Difference in ${\tt pmdec\_hg}$ from the catalog-computed value: $\xi_{\delta,HG}$ from Table \ref{tab:hgca_2} \\
${\tt crosscal\_pmra\_hip}$ & mas\,yr$^{-1}$ & Difference in ${\tt pmra\_hip}$ from the catalog-computed value: $\xi_{\alpha*,H}$ from Table \ref{tab:hgca_1} \\
${\tt crosscal\_pmdec\_hip}$ & mas\,yr$^{-1}$ & Difference in ${\tt pmra\_hip}$ from the catalog-computed value: $\xi_{\delta,H}$ from Table \ref{tab:hgca_1} \\
${\tt nonlinear\_dpmra}$ & mas\,yr$^{-1}$ & Correction to ${\tt pmra\_hg}$ from projecting linear motion onto the celestial sphere: $\gamma_{\alpha*}$ from Table \ref{tab:hgca_2} \\
${\tt nonlinear\_dpmdec}$ & mas\,yr$^{-1}$ & Correction to ${\tt pmdec\_hg}$ from projecting linear motion onto the celestial sphere: $\gamma_{\delta}$ from Table \ref{tab:hgca_2}
\enddata
\end{deluxetable*}

\subsection{User Guidelines}

The Hipparcos-Gaia Catalog of Accelerations is intended to identify targets to search for substellar and dark companions, and to derive dynamical masses.  The covariance matrices should, in most cases, be reliable enough for orbit fitting; the user can (and should) subtract proper motions and add their covariance matrices for these purposes.  This may not be true for stars with especially large {\it Gaia} uncertainties.  The catalog is {\it not} intended to statistically constrain the distribution of binary properties and binary orbital parameters.  We urge users not to apply it to such an analysis, except on a star-by-star basis where the companions are known and can have their orbits fit.  

The catalog does have a small number of spurious matches and stars with poor astrometry.  Any analysis relying on the statistical properties of the catalog must treat these cases very carefully.  Figure \ref{fig:final_uncertainties} suggests that a small fraction of the {\it Gaia} uncertainties remain incorrect, especially for targets with already large uncertainties ($\gtrsim$0.7~mas\,yr$^{-1}$ after the inflation we apply).  
Finally, many of the accelerating stars are binaries with modest brightness ratios.  These binaries can cause problems particularly with the lower-resolution {\it Hipparcos} data, and should also be treated with caution.

Again, we stress that for a star identified by other means (like a radial velocity trend) and without a close binary or neighbor of comparable brightness, the covariance matrices are unlikely to be problematic.  This catalog is intended for use on a case-by-case, star-by-star basis; under those circumstances it should be a reasonably robust source of astrometric accelerations.  

\cite{Brandt+Dupuy+Bowler_2018} provide some examples of the use of this catalog in fitting orbits and obtaining dynamical masses.  The catalog contains three effectively independent proper motion measurements and covariance matrices.  These measurements may be used straightforwardedly in a $\chi^2$ likelihood framework for constraining orbits and masses.

\section{Conclusions} \label{sec:conclusions}

In this paper, we have presented a catalog cross-calibrating the {\it Hipparcos} and {\it Gaia} astrometry to enable fitting of astrometric accelerations.  Offsets in reference frame and discrepancies between the formal and actual uncertainties are expected and, in some cases, calculated by the {\it Hipparcos} and {\it Gaia} teams; they must be accounted for before fitting accelerations.  Our cross-calibration takes this one step further to compute a locally variable cross-calibration. 

We provide three sets of proper motions, one for {\it Hipparcos}, one for {\it Gaia}, and one for the scaled positional difference.  We also provide covariance matrices for each.  The cross-covariances between these sets of proper motions should be very nearly zero; the proper motions may be considered to be independent.  Together with star-by-star observational epochs, this provides data in a way that is straightforward to implement in orbit fitting codes.

We have performed a global fit to nine parameters to place the {\it Hipparcos} and {\it Gaia} data sets on a common reference frame, including a variable weighting of the two {\it Hipparcos} reductions.  We have then shown that a locally variable fit, with the smoothing parameters optimized using a Gaussian process regression, is decisively favored in a cross-validation data set over the global best-fit parameters.  Our analysis shows that, of the two {\it Hipparcos} reductions, the best is actually a linear combination of the two.  Each instrument also requires inflation of its uncertainties, albeit of a different form for {\it Hipparcos} and {\it Gaia}.  Our final catalog shows Gaussian residuals in proper motion differences.  After normalizing these by our calibrated errors, the distributions closely match Gaussians of unit variance.

The catalog is nearly complete for the {\it Hipparcos} sample, containing {\it Gaia} DR2 matches for more than 97\% of its stars.  A few hundred of these are spurious and/or have poor {\it Hipparcos} reductions; the user is responsible for vetting the basic data quality star-by-star.  For this reason, because of heavy tails in the {\it Gaia} proper motion residuals for low-precision stars, and because of contamination by blended light from binaries, the catalog is probably not suitable for constraining the distribution of accelerations.  It is instead intended for orbit fitting and for searching for accelerating stars.  We have compiled the catalog with an eye toward completeness, preferring to include a few bad entries over the exclusion of a few valid ones.

\acknowledgments{TDB thanks Scott Tremaine, Daniel Michalik, Alcione Mora, and Anthony Brown for input on a draft of this paper, and Ron Drimmel for showing him how to query the {\it Gaia} database.  He thanks Daniel Michalik, Trent Dupuy, Brendan Bowler, and Jackie Faherty for inspiration to compile this catalog, and Trent in particular for helping to test it as an orbit fitting tool.  TDB thanks an anonymous referee for a careful reading and helpful suggestions.  This work has made use of data from the European Space Agency (ESA) mission {\it Gaia} (https://www.cosmos.esa.int/gaia), processed by the Gaia Data Processing and Analysis Consortium (DPAC, https://www.cosmos.esa.int/web/gaia/dpac/consortium). Funding for the DPAC has been provided by national institutions, in particular the institutions participating in the {\it Gaia} Multilateral Agreement.  TDB gratefully acknowledges support from the Heising-Simons foundation.}

\bibliographystyle{apj_eprint}
\bibliography{refs.bib}

\begin{thebibliography}{}

\bibitem[\protect\citeauthoryear{{Brandt}, {Dupuy}, \& {Bowler}}{{Brandt}
  et~al.}{2018}]{Brandt+Dupuy+Bowler_2018}
{Brandt}, T.~D., {Dupuy}, T.,  \& {Bowler}, B.~P. 2018, submitted

\bibitem[\protect\citeauthoryear{{Calissendorff} \& {Janson}}{{Calissendorff}
  \& {Janson}}{2018}]{Calissendorff+Janson_2018}
{Calissendorff}, P.,  \& {Janson}, M. 2018, ArXiv e-prints, 1806.07899

\bibitem[\protect\citeauthoryear{{ESA}}{{ESA}}{1997}]{ESA_1997}
{ESA}, ed. 1997, ESA Special Publication, Vol. 1200, {The HIPPARCOS and TYCHO
  catalogues. Astrometric and photometric star catalogues derived from the ESA
  HIPPARCOS Space Astrometry Mission}

\bibitem[\protect\citeauthoryear{{Fey} et~al.}{{Fey}
  et~al.}{2015}]{Fey+Gordon+Jacobs+etal_2015}
{Fey}, A.~L., {Gordon}, D., {Jacobs}, C.~S., et~al. 2015, \aj, 150, 58

\bibitem[\protect\citeauthoryear{{Gagn{\'e}} \& {Faherty}}{{Gagn{\'e}} \&
  {Faherty}}{2018}]{Gagne+Faherty_2018}
{Gagn{\'e}}, J.,  \& {Faherty}, J.~K. 2018, \apj, 862, 138

\bibitem[\protect\citeauthoryear{{Gaia Collaboration} et~al.}{{Gaia
  Collaboration} et~al.}{2018a}]{Gaia_General_2018}
{Gaia Collaboration}, {Brown}, A.~G.~A., {Vallenari}, A., et~al. 2018a, \aap,
  616, A1

\bibitem[\protect\citeauthoryear{{Gaia Collaboration} et~al.}{{Gaia
  Collaboration} et~al.}{2018b}]{Gaia_Clusters_2018}
{Gaia Collaboration}, {Helmi}, A., {van Leeuwen}, F., et~al. 2018b, \aap, 616,
  A12

\bibitem[\protect\citeauthoryear{{Gaia Collaboration} et~al.}{{Gaia
  Collaboration} et~al.}{2016}]{Gaia_General_2016}
{Gaia Collaboration}, {Prusti}, T., {de Bruijne}, J.~H.~J., et~al. 2016, \aap,
  595, A1

\bibitem[\protect\citeauthoryear{Gneiting}{Gneiting}{2013}]{Gneiting_2013}
Gneiting, T. 2013, Bernoulli, 19, 1327

\bibitem[\protect\citeauthoryear{{Ivezi{\'c}} et~al.}{{Ivezi{\'c}}
  et~al.}{2014}]{Ivezic+Connolly+Vanderplas+etal_2014}
{Ivezi{\'c}}, {\v Z}., {Connolly}, A., {Vanderplas}, J.,  \& {Gray}, A. 2014,
  Statistics, Data Mining and Machine Learning in Astronomy (Princeton
  University Press)

\bibitem[\protect\citeauthoryear{{Koppelman}, {Helmi}, \&
  {Veljanoski}}{{Koppelman} et~al.}{2018}]{Koppelman+Helmi+Veljanoski_2018}
{Koppelman}, H., {Helmi}, A.,  \& {Veljanoski}, J. 2018, \apjl, 860, L11

\bibitem[\protect\citeauthoryear{{Lindegren} et~al.}{{Lindegren}
  et~al.}{2018}]{Gaia_Astrometry_2018}
{Lindegren}, L., {Hernandez}, J., {Bombrun}, A., et~al. 2018, arxiv, 1804.09366

\bibitem[\protect\citeauthoryear{{Lindegren} et~al.}{{Lindegren}
  et~al.}{2016}]{TGAS_Astrometry_2016}
{Lindegren}, L., {Lammers}, U., {Bastian}, U., et~al. 2016, \aap, 595, A4

\bibitem[\protect\citeauthoryear{{Lindegren} et~al.}{{Lindegren}
  et~al.}{2012}]{Lindegren+Lammers+Hobbs+etal_2012}
{Lindegren}, L., {Lammers}, U., {Hobbs}, D., et~al. 2012, \aap, 538, A78

\bibitem[\protect\citeauthoryear{{Ma} et~al.}{{Ma}
  et~al.}{1998}]{Ma+Arias+Eubanks+etal_1998}
{Ma}, C., {Arias}, E.~F., {Eubanks}, T.~M., et~al. 1998, \aj, 116, 516

\bibitem[\protect\citeauthoryear{{Malo} et~al.}{{Malo}
  et~al.}{2013}]{Malo+Doyon+Lafreniere_2013}
{Malo}, L., {Doyon}, R., {Lafreni{\`e}re}, D., et~al. 2013, \apj, 762, 88

\bibitem[\protect\citeauthoryear{{Marrese} et~al.}{{Marrese}
  et~al.}{2018}]{Marrese+Marinoni+Fabrizio+etal_2018}
{Marrese}, P.~M., {Marinoni}, S., {Fabrizio}, M.,  \& {Altavilla}, G. 2018,
  ArXiv e-prints, 1808.09151

\bibitem[\protect\citeauthoryear{{Michalik}, {Lindegren}, \&
  {Hobbs}}{{Michalik} et~al.}{2015}]{Michalik+Lindegren+Hobbs_2015}
{Michalik}, D., {Lindegren}, L.,  \& {Hobbs}, D. 2015, \aap, 574, A115

\bibitem[\protect\citeauthoryear{{Michalik} et~al.}{{Michalik}
  et~al.}{2014}]{Michalik+Lindegren+Hobbs+etal_2014}
{Michalik}, D., {Lindegren}, L., {Hobbs}, D.,  \& {Lammers}, U. 2014, \aap,
  571, A85

\bibitem[\protect\citeauthoryear{{Mignard} et~al.}{{Mignard}
  et~al.}{2018}]{Gaia_ICRF_2018}
{Mignard}, F., {Klioner}, S., {Lindegren}, L., et~al. 2018, ArXiv e-prints,
  1804.09377

\bibitem[\protect\citeauthoryear{{Price-Whelan} \& {Johnston}}{{Price-Whelan}
  \& {Johnston}}{2013}]{Price-Whelan+Johnston_2013}
{Price-Whelan}, A.~M.,  \& {Johnston}, K.~V. 2013, \apjl, 778, L12

\bibitem[\protect\citeauthoryear{{Sanderson}, {Hartke}, \& {Helmi}}{{Sanderson}
  et~al.}{2017}]{Sanderson+Hartke+Helmi_2017}
{Sanderson}, R.~E., {Hartke}, J.,  \& {Helmi}, A. 2017, \apj, 836, 234

\bibitem[\protect\citeauthoryear{{Simon}}{{Simon}}{2018}]{Simon_2018}
{Simon}, J.~D. 2018, \apj, 863, 89

\bibitem[\protect\citeauthoryear{{Snellen} \& {Brown}}{{Snellen} \&
  {Brown}}{2018}]{Snellen+Brown_2018}
{Snellen}, I.~A.~G.,  \& {Brown}, A.~G.~A. 2018, Nature Astronomy, 1808.06257

\bibitem[\protect\citeauthoryear{{van Leeuwen}}{{van
  Leeuwen}}{2007}]{vanLeeuwen_2007}
{van Leeuwen}, F. 2007, \aap, 474, 653

\bibitem[\protect\citeauthoryear{{Zuckerman}, {Song}, \& {Bessell}}{{Zuckerman}
  et~al.}{2004}]{Zuckerman+Song+Bessell_2004}
{Zuckerman}, B., {Song}, I.,  \& {Bessell}, M.~S. 2004, \apjl, 613, L65

\bibitem[\protect\citeauthoryear{{Zuckerman} et~al.}{{Zuckerman}
  et~al.}{2001}]{Zuckerman+Song_2001}
{Zuckerman}, B., {Song}, I., {Bessell}, M.~S.,  \& {Webb}, R.~A. 2001, \apjl,
  562, L87

\end{thebibliography}

\end{document}